\documentclass[
a4paper,
aps,
secnumarabic,
prd, 
preprint,
showpacs,
showkeys,
showbib,
nofootinbib,
longbibliography,
]{revtex4-1}
\usepackage{amsmath, amsthm, amssymb,latexsym,mathrsfs,dsfont}    
\usepackage{float}
\usepackage{graphicx}   
\usepackage{geometry}
\usepackage{verbatim}   
\usepackage{listings}
\usepackage{xcolor}
\usepackage{color}   

\usepackage[caption=false]{subfig} 
\usepackage{braket}
\usepackage{cancel}
\usepackage{slashed}
\usepackage{bookmark,bbm}
\newcommand{\vect}[1]{\boldsymbol{#1}} 
\usepackage{cancel}
\usepackage{simplewick}
\graphicspath{{./Figs/}}
\newcommand{\eq}[1]{\begin{equation}\begin{aligned}#1\end{aligned}\end{equation}}
\newcommand{\ee}{\mathrm{e}}
\newcommand{\ii}{\mathrm{i}} 
\newcommand{\dd}{\mathrm{d}}

\def\dbar{{\mathrm{\mathchar'26\mkern-10mu d}}} 
\DeclareMathOperator{\tr}{tr}

\hypersetup{pdftitle={main.pdf},hidelinks
}
\begin{document}


\title{Octet and decuplet contribution to the proton self energy}

\author{Hazhar Ghaderi}
\email{hazhar.ghaderi@physics.uu.se}
\affiliation{Department of Physics and Astronomy, Uppsala University, Box 516, SE-751 20 Uppsala, Sweden}
\date{July 17, 2018}

\begin{abstract} 
Within the hadronic language of Chiral Perturbation Theory we present the full leading-order octet-baryon--meson and decuplet-baryon--meson contribution to the proton self energy and thus to its wave function renormalization factor $Z$. By Fock-expanding the physical proton state into its bare and hadron-cloud part, we show how each individual baryon-meson probability depend on the average momenta of the particles in the fluctuation. We present how the results depend on the choice of the form factor involved in the regularization (Gaussian or Besselian) and how they depend on the cut-off parameter. We also show how the results vary with respect to a variation of the decuplet coupling constant $h_A$. The momentum distributions of the fluctuations are given and the fluctuations' relative probabilities are presented. 

We show that for reasonable values of the cut-off parameter, the Delta-pion fluctuation is of the same strength as the nucleon-pion fluctuation.   
\end{abstract}



\maketitle
\thispagestyle{empty}
\null\newpage\thispagestyle{empty}
\clearpage\setcounter{page}{1}
\section{Introduction}\label{Sec:1}
Many of the properties which we take for granted from a classical point of view, such as e.g.\ the electric charge or even the mass of a microscopic particle are not that simple when one takes a deeper look. Not long after the foundations of Quantum Mechanics (QM) were laid, the search for a theory consistently describing creation of particle-antiparticle pairs was pursued and highly sought for. This was enhanced by the experimental detection of antiparticles \cite{Anderson:1932zz, *Anderson:1933mb}. Apart from the intermediate years when S-matrix\footnote{See \cite{Weinberg:1996kw} for a shorter discussion on QFT and the S-matrix. } methods were perhaps favorable, the main contestant and candidate was the theory of second quantization or what is now more commonly called Quantum Field Theory (QFT). The problem now was that in such a theory the number of degrees of freedom (DOF) were infinite and the theory often gave unsatisfying results, e.g.\ an infinite term plus some terms that seemed to look fine. It took its time but for the QFT of Quantum Electrodynamics (QED) these infinities were finally understood and dealt with via \emph{renormalization} techniques (see e.g.\ \cite{PhysRev.75.486} and references therein). These techniques include some type of renormalization that naturally introduces an energy dependence for the `constants' of the theory, such as the electric charge $e$ (of an electron say) and effectively replaces them by functions $e\rightarrow e(Q)$ where $Q$ is the energy-momentum transfer in the reaction. Pictorially one could imagine the electron being surrounded by particle-antiparticle pairs effectively screening the \emph{bare} charge of the electron \cite{Peskin:257493}. For instance, at energies on the order of the mass of the $Z$ and the $W$ bosons, the fine structure coupling $\alpha\equiv e^2/4\pi$ has a value of approximately 1/128 as compared to $\alpha\approx1/137$ at much smaller energy scales. Thus, these fluctuations and other vacuum polarization effects have impact on what one would call the electric charge of the electron. 

The particle-antiparticle pairs have the quantum numbers of the object that one studies. They can be understood to be energetically allowed via Heisenberg's uncertainty relation. Other than mass suppression, there is no reason why similar effects as the ones described above wouldn't appear elsewhere. We are here interested in the case of the proton (or nucleon in general) and the hadronic parts of its wavefunction. By this we mean to calculate the coefficients $\alpha_{ij}$ of its Fock expansion (see e.g.\ \cite{Holtmann:1996be, Edin:1998dz, Alwall:2005xd, Peng:2014eza, Kofler:2017uzq} for other/similar applications)
\eq{\label{E: FockExp}
	\ket{P} = \sqrt{Z}\ket{P}_\text{bare}  + \alpha_{n^0\pi^+}
	\ket{n^0\pi^+}
	+ \alpha_{P\pi^0}\ket{P\pi^0}
	+ \alpha_{\Delta^{++}\pi^-}\ket{\Delta^{++}\pi^-}
	+
	\cdots
}
where $\ket{P}$ is the physical proton state and the coefficients on the right hand side are those for the bare proton $(\sqrt{Z})$, the neutron-$\pi^+$ fluctuation $(\alpha_{n^0\pi^+})$ and so on. Using the leading-order Lagrangian of Chiral Perturbation Theory $(\chi\text{PT})$ \cite{Scherer:2002tk} we will include all the octet-baryon--meson and decuplet-baryon--meson pairs in the expansion (\ref{E: FockExp}). 

We start by introducing the relevant $\chi\text{PT}$ Lagrangian in Section \ref{Sec: theLagrangian}. Then in Section \ref{Sec: Z} we will write down an expression for the wave function renormalization constant $Z$. In the same section we will derive and present an analytic expression for the momentum distributions of the hadronic fluctuations. We will then show how the integrated momentum distributions are related to the baryon-meson probabilities $|\alpha_{BM}|^2$ of Equation (\ref{E: FockExp}). 

By construction $\chi\text{PT}$ is an effective\footnote{We refer the reader to \cite{doi:10.1146/annurev.ns.43.120193.001233, Georgi:1991mr,Leupold:2012qn, Manohar:1997qy, Bauer:2002uv} and the references therein for various different examples of effective theories. } (low energy limit) theory of Quantum Chromodynamics (QCD) where the DOF are hadrons and not quarks and gluons which are the microscopic DOF of QCD. Perturbative Quantum Chromodynamics (pQCD) breaks down at energy-momentum transfers $Q\lesssim 1~\text{GeV}$ where the appropriate description instead is the hadron language. Vice versa one should not apply $\chi\text{PT}$ to problems involving energy scales as large as $Q\approx 1~\text{GeV}$ where the constituents of the hadrons (quarks and gluons) become the relevant DOF. We take this into account in our calculations by introducing a cut-off via a form factor which we discuss in Section \ref{Sec: FF}. 
In Section \ref{Sec: Results} we present our results and we conclude in Section \ref{Sec: Conclusions}.  
\section{The Lagrangian}\label{Sec: theLagrangian}
For the case at hand, the Rarita-Schwinger Lagrangian is sufficient to describe the propagation of massive spin-3/2 particles. For a decuplet-hadron of mass $m_D$, the Rarita-Schwinger Lagrangian is given by \cite{Rarita41}
\eq{
	\mathcal{L}_{\text{RS}}  = \bar{\psi}_\mu \Lambda^{\mu\nu}_{\text{RS}}\psi_\nu, 
}
where 
\eq{
	\Lambda^{\mu\nu}_{\text{RS}} = g^{\mu\nu}(\slashed{p}-m_D)-(\gamma^\mu p^\nu+p^\mu \gamma^\nu)+\gamma^\mu(\slashed{p}+m_D)\gamma^\nu = \frac{\ii}{2}\{\sigma^{\mu\nu},(\slashed{p}-m_D)\}  
}
with $\slashed{p} = \ii \slashed{\partial}$. 
To find the propagator of the Rarita-Schwinger field, one solves $\Lambda^\text{RS}_{\mu\rho}G_\text{RS}^{\rho\nu} = g{_\mu^\nu}$ for $G_\text{RS}^{\rho\nu} $ and obtains 
\eq{\label{E: RSPropagator}
	G_\text{RS}^{\mu\nu} = -\frac{\slashed{p}+m_D}{p^2-m_D^2}\left[g^{\mu\nu}-\frac{1}{3}\gamma^\mu\gamma^\nu-\frac{2p^\mu p^\nu}{3m_D^2}+\frac{p^\mu\gamma^\nu-p^\nu\gamma^\mu  }{3m_D}\right]. 
} 
The tensor in square brackets is called the Rarita-Schwinger tensor, it also propagates the spin-1/2 parts of the representation (see e.g.\ \cite{Haberzettl:1998rw}) which can be problematic and unphysical. This is related to the number of DOF of an \emph{elementary} vector spinor $\psi^\mu$. A spin-3/2 state should have 8 DOF while the aforementioned has 16 DOF. With this in mind, the Lagrangian for the free theory is constructed to give rise to not only the equations of motion but also to constraints that reduce the number of DOF. However once interactions are introduced other issues which also have been shown to be related to the number of DOF \cite{Cox:1989hp} arise again. We refer to the introduction of \cite{Jahnke:2006nj} for a longer discussion on this but for our case what it all boils down to is that as long as one stays within the range of applicability of the effective (hadron) theory, one should be fine by simply subscribing to the Pascalutsa prescription \cite{Pascalutsa:1999zz} (as discussed in the following). 

The Lagrangian describing the interaction of the Goldstone bosons with octet and decuplet baryons is given by (see \cite{Granados:2017cib} and the references therein)

\eq{\label{E: DNpi}
	\mathcal{L}_{8+10} = \ii&\tr (\bar{B}\gamma_\mu D^\mu B) +\frac{D}{2}\tr(\bar{B} \gamma^\mu \gamma_5 \{u_\mu,B\})+\frac{F}{2}\tr (\bar{B}\gamma^\mu\gamma_5[u_\mu,B])
	\\
	+&\frac{1}{2\sqrt{2}}h_A\epsilon_{ade}g_{\mu\nu}\left(\bar{T}^\mu_{abc}u^\nu_{bd} B_{ce}+\bar{B}_{ec}u^\nu_{db} T_{abc}^\mu\right).  
}
As it stands, it induces unphysical contact interactions from the spurious spin-1/2 admixture. To cure this we make the Pascalutsa substitution in the Lagrangian by letting 
\eq{\label{E: Pascalutsa}
	T^\mu\rightarrow -\frac{1}{m_R}\epsilon^{\nu\mu\alpha\beta}\gamma_5\gamma_\nu \partial_\alpha T_\beta 
}
where $m_R = m_\Delta, m_{\Sigma^{*}}$ denotes the resonance mass.  What appears in the coupling after the substitution is the ratio $h_A/m_R$ and since the value of $h_A$ has some uncertainty in it (see below) one can in practice use $m_R=m_\Delta$ and check how a variation in $h_A$ affects the results. 
Note that this substitution induces an explicit flavor breaking but these effects are beyond leading order. 
In Equation (\ref{E: DNpi}) $B_{ab}$ is the entry in the $a$th row, $b$th column of the matrix representing the octet baryons  
\eq{
	B = 
	\begin{pmatrix}
    \frac{1}{\sqrt{2}}\Sigma^0+\frac{1}{\sqrt{6}}\Lambda       & \Sigma^+ & P \\
    \Sigma^-       & -\frac{1}{\sqrt{2}}\Sigma^0+\frac{1}{\sqrt{6}}\Lambda & n \\
   \Xi^-       & \Xi^0 & -\frac{2}{\sqrt{6}}\Lambda
\end{pmatrix}. 
}
The Goldstone bosons are contained in 
\eq{
	\Phi = 
	\begin{pmatrix}
    \pi^0+\frac{1}{\sqrt{3}}\eta      & \sqrt{2}\pi^+ & \sqrt{2}K^+ \\
    \sqrt{2}\pi^-       & -\pi^0+\frac{1}{\sqrt{3}}\eta& \sqrt{2}K^0 \\
   \sqrt{2}K^-       & \sqrt{2}\bar{K}^0 & -\frac{2}{\sqrt{3}}\eta
\end{pmatrix} 
}
and $u_\mu$ is given by 
\eq{
	u_\mu = \ii u^\dagger(\nabla_\mu U)u^\dagger  = u_\mu^\dagger
}
where 
\eq{
	u^2 = U = \exp(\ii \Phi /F_\pi). 
}
The pion decay constant is denoted by $F_\pi = 92.4\text{ MeV}$ and the chirally covariant derivatives are defined by 
\eq{
	D^\mu B = \partial^\mu B+ [\Gamma^\mu, B]
}
and 
\eq{
	\nabla_\mu = \partial_\mu U -\ii(v_\mu+a_\mu)U+\ii U(v_\mu-a_\mu), 
}
with 
\eq{
	\Gamma_\mu  = \frac{1}{2} \left[ u^\dagger(\partial_\mu-\ii(v_\mu+a_\mu)) u+u(\partial_\mu-\ii(v_\mu-a_\mu))u^\dagger \right]
} 
where $v_\mu, a_\mu$ are external sources. In our case, there are no external sources, thus $v=a =0$. Finally, the decuplet is represented by a totally symmetric flavor tensor 
\eq{\label{E: DecupletPart}
	&T^{111} = \Delta^{++},T^{112} = \frac{1}{\sqrt{3}}\Delta^{+},T^{122} = \frac{1}{\sqrt{3}}\Delta^{0},T^{222} = \Delta^{-}, \\
	&T^{113} = \frac{1}{\sqrt{3}}\Sigma^{*+}, T^{123} = \frac{1}{\sqrt{6}}\Sigma^{*0}, T^{223} = \frac{1}{\sqrt{3}}\Sigma^{*-}, \\
	&T^{133} = \frac{1}{\sqrt{3}}\Xi^{*0}, T^{233} = \frac{1}{\sqrt{3}}\Xi^{*-}, 
	\\
	&T^{333} = \Omega. 
}
For us only the first two rows of Equation (\ref{E: DecupletPart}) are allowed in the interactions due to conservation of quantum numbers. 
The coupling $h_A$ in Equation (\ref{E: DNpi}) can be determined by matching to experimental data from the partial decay width $\Delta\rightarrow N\pi $ or from $\Sigma^{*}\rightarrow \Lambda\pi$ to be 
\eq{\label{E: hupperlower}
	&h_A^{\Delta\rightarrow N\pi} = 2.88, \\
	&h_A^{\Sigma^{*}\rightarrow \Lambda\pi} = 2.4.  
}
In the large-$N_C$ limit, one can also relate it to another for us relevant coupling \cite{Dashen:1993as, Pascalutsa:2005nd}
\eq{
	 h_A^\text{large-$N_C$} = \frac{3}{\sqrt{2}}g_A = 2.67 
}
where $g_A = 1.26 = D+F$. The couplings $D$ and $F$ which show up in the octet part of the Lagrangian (\ref{E: DNpi}) also satisfy $F-D = -0.34$ \cite{Cabibbo:2003cu}. They may vary independently $\pm 10\%$, but they must sum to $g_A$ which has been measured very precisely in e.g.\ neutron $\beta$-decay, pion-nucleon scattering and long-range nucleon-nucleon scattering. The value for $g_A$ obtained from these experiments agree\footnote{The disagreement comes from  electromagnetic effects and quark-mass differences. } on a level of $1\%$ \cite{Patrignani:2016xqp}. We will here choose to work with the fixed values of $D=0.8$ and $F = 0.46$ since it turns out that an independent variation does not affect the results by much, since it is anyway their sum that enters the most probable fluctuations (see Table \ref{Table: Coupling}). For the masses of the hadrons we use the values collected in Table 
\ref{Table: masses}, where we have neglected any mass-differences within each isospin multiplet. 
\begin{table}
\caption{Coupling constants for the decuplet-baryon--meson and octet-baryon--meson pairs that couple to an initial state proton. The relative strength of each contribution is also shown. 
}\label{Table: Coupling}
\resizebox{1.0\textwidth}{!}{
\renewcommand{\arraystretch}{1.1}
\begin{tabular}{ |c ||c |c |c |c |c |c |  }
 \hline
 \multicolumn{7}{|c|}{Decuplet-baryon--meson} \\
 \hline
$DM$ &   $\Delta^{++}\pi^-$&$\Delta^{+}\pi^0$&$\Delta^{0}\pi^+$&$\Sigma^{*+}K^0$&$\Sigma^{*0}K^+$&\\
 $g_{DM}$   &     $\frac{h_A}{2m_R F_\pi}$& $\frac{-h_A}{\sqrt{6}m_R F_\pi}$& $\frac{-h_A}{2\sqrt{3}m_R F_\pi}$& $\frac{h_A}{2\sqrt{3}m_R F_\pi}$& $\frac{-h_A}{2\sqrt{6}m_R F_\pi}$& \\
 $\left|{g_{DM}}/{g_{\Delta^{++}\pi^-}}\right|^2 $&  1  &  0.67 & 0.33 & 0.33 & 0.17 &   \\
 \hline
 \hline
\multicolumn{7}{|c|}{Octet-baryon--meson} \\
 \hline
$OM$ & $n\pi^+$ &$P\pi^0$&$\Lambda K^+ $ & $\Sigma^+K^0$&$\Sigma^{0}K^+$&$P\eta$\\
 $g_{OM}$   & $-\frac{D+F}{\sqrt{2}F_\pi}$    &$-\frac{D+F}{2F_\pi}$& $\frac{D+3F}{2\sqrt{3}F_\pi}$& $-\frac{D-F}{\sqrt{2}F_\pi}$& $-\frac{D-F}{2F_\pi}$& $\frac{D-3F}{2\sqrt{3}F_\pi}$
 \\
 $\left|{g_{OM}}/{g_{n\pi^+}}\right|^2$ &  1  & 0.5   & 0.5 & 0.08 & 0.04 & 0.03 
\\
\hline
\end{tabular}
}
\end{table} 
\begin{table}
\caption{The values for the hadron masses \cite{Patrignani:2016xqp}. }\label{Table: masses}
\begin{tabular}{|l*{7}{|c}|r|}
\hline
Hadron mass ~~& $m_{\Sigma^{*}}$ & $m_\Delta$ & $m_{\Sigma^{}}$ & $m_\Lambda$ & $m_N $  & $m_\eta$ & $m_K$ & $m_\pi$ \\
\hline
Value [MeV] & 1382 & 1232 & 1189 & 1115 & 938 & 547 & 493 & 139 \\
\hline
\end{tabular} 
\end{table}

Making the Pascalutsa substitution (\ref{E: Pascalutsa}) in the  Lagrangian (\ref{E: DNpi}) and writing out the explicit terms that couple to an initial and final state proton $P$, one obtains for the decuplet part of the Lagrangian 
\eq{\label{E: decLag}
	&\mathcal{L}^{PDM} = ~\frac{h_A 
	\varepsilon^{\rho\mu\alpha\beta}
	}{2m_R F_\pi}
	\Bigg[ 
	\sqrt{\frac{1}{3}}(\partial_\alpha \bar{\Sigma}^{*+}_\beta)
	\gamma_5\gamma_\rho(\partial_\mu \bar{K}^0)
	- \sqrt{\frac{1}{6}}(\partial_\alpha \bar{\Sigma}^{*0}_\beta)
	\gamma_5\gamma_\rho(\partial_\mu K^-)
	\\
	&
	+ 
	(\partial_\alpha \bar{\Delta}^{++}_\beta)
	\gamma_5\gamma_\rho(\partial_\mu\pi^+)
	- \sqrt{\frac{2}{3}}(\partial_\alpha \bar{\Delta}^{+}_\beta)
	\gamma_5\gamma_\rho(\partial_\mu\pi^0)
	- \sqrt{\frac{1}{3}}(\partial_\alpha \bar{\Delta}^{0}_\beta)
	\gamma_5\gamma_\rho(\partial_\mu\pi^-) 
	\Bigg]P +\!\text{hc.}  
}
Doing the same for the octet part of the Lagrangian 
\eq{\label{E: octLag}
	\mathcal{L}^{POM} = \Bigg[&-\frac{D+F}{\sqrt{2} F_{\pi }}\bar{n} 
	\gamma^\mu\gamma_5 (\partial_\mu \pi^-)  
	-\frac{D+F}{2 F_{\pi }} \bar{P}
	\gamma^\mu\gamma_5 (\partial_\mu   \pi^0) 
	+
	\frac{D-3 F}{2\sqrt{3} F_{\pi }}\bar{P}
	\gamma^\mu\gamma_5 (\partial_\mu  \eta) 
	\\
	&-\frac{D-F}
	{2 F_{\pi }}\bar{\Sigma }^0 
	\gamma^\mu\gamma_5 (\partial_\mu K^-)
	-\frac{ D-F}
	{\sqrt{2} F_{\pi }}\bar{\Sigma}^+ 
	\gamma^\mu\gamma_5 (\partial_\mu  \bar{K}^0) 
	\\
	&
	+  
	\frac{ D+3 F}{2\sqrt{3} F_{\pi }}\bar{\Lambda}
	\gamma^\mu\gamma_5 (\partial_\mu K^-)  
	\Bigg] P +\text{hc.}
}
we identify all the relevant couplings, which we have collected in Table \ref{Table: Coupling}. We have also written the strength of each contribution relative to the largest coupling within each multiplet. These numbers are the respective strengths of each fluctuation probability that contributes to the total fluctuation probability $1-Z$, a subject which we now will devote our attention to. 

\section{The Wave Function Renormalization}\label{Sec: Z}
Let $\ket{\Omega}$ denote the ground state of the interacting theory. Then the two-point function of Dirac fields has the following structure \cite{Peskin:257493}
\eq{
	\int\!\dd^4x\,\ee^{\ii p\cdot x}\braket{\Omega|T \psi(x)
	\bar{\psi}(0)|\Omega}
	&=
	\sum_s\frac{\ii Z u^s(p)\bar{u}^s(p)}{p^2-m^2+\ii\epsilon}
	+\cdots
	\\
	&=
	\frac{\ii Z(\slashed{p}+m)}{p^2-m^2+\ii\epsilon}+\cdots
}
where the dots stands for contribution from the multiparticle branch cut. The rescaling constant $Z$ is the probability for the quantum field to create or annihilate an exact one-particle eigenstate of the hamiltonian  
\eq{
	\braket{\Omega|\psi(0)|p,s} = \sqrt{Z}u^s(p).  
}

\begin{figure}[t]
\begin{center}
\includegraphics[width = 0.8\linewidth]{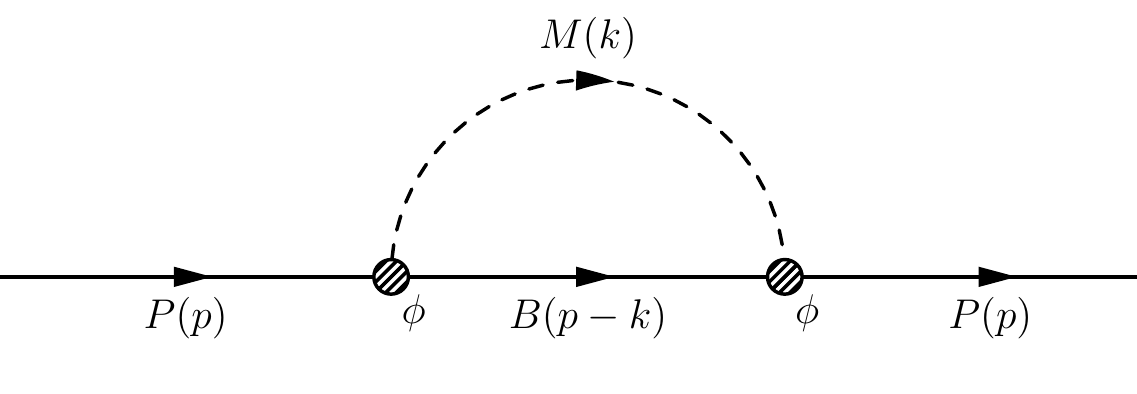}
\caption{The Feynman diagram for the baryon-meson contribution to the proton self-energy where the baryon belongs to the octet or the decuplet. The blob denoted by $\phi$ represents the form factor. }
\label{Fig: NpiSelf_Energy}
\end{center}
\end{figure}

The free-field part of the Lagrangian (\ref{E: DNpi}) yields the lowest-order Feynman propagator of the nucleon 
\eq{\label{E: FeynmanProp}
	\ii S_F(p) = \frac{\ii}{\slashed{p}-\overset{\circ}{m}_N+\ii \epsilon}, 
}
where $\overset{\circ}{m}_N$ denotes the bare mass of the nucleon. 
The Feynman propagator (\ref{E: FeynmanProp}) is modified by the self energy $\Sigma(p)$ resulting in the full \emph{unrenormalized} propagator
\eq{
	\ii S_F(p)\rightarrow \frac{\ii}{\slashed{p}-\overset{\circ}{m}_N+\ii \epsilon}+\frac{\ii}{\slashed{p}-\overset{\circ}{m}_N+\ii \epsilon}[-\ii\Sigma(p)]\frac{\ii}{\slashed{p}-\overset{\circ}{m}_N+\ii \epsilon}
	+\cdots 
	=\ii S(p). 
}
Thus, summing the geometric series we can write 
\eq{\label{E: UnReProp}
	\ii S(p) = \frac{\ii}{\slashed{p}-\overset{\circ}{m}_N-\Sigma(p)+\ii \epsilon}. 
}
We can write down the most general expression for the self energy (without any external sources) as the sum of a scalar and a vector part 
\eq{\label{E: scalarvector}
	\Sigma(p) = \Sigma_s(p^2)+\slashed{p} F(p^2), 
}
where the two Lorentz invariant functions $\Sigma_s(p^2)$ and $F(p^2)$ can be projected out by 
\eq{\label{E: projection}
	F(p^2)\Big\rvert_{p^2=m_N^2} &=    \frac{1}{4p^2}\tr(\slashed{p}\Sigma)\Bigg\rvert_{p^2=m_N^2}
	\\
	\Sigma_s(p^2) &= \frac{1}{4}\tr(\Sigma)\Bigg\rvert_{p^2=m_N^2}. 
}
Using the pole mass definition 
\eq{
	m_N-\overset{\circ}{m}_N-m_NF(m_N^2)-\Sigma_s(m_N^2)=0 
}
and the definition of the renormalized propagator 
\eq{
	S(p) = Z S_R(p)
}
having a pole of $\text{residue}=1$ at $\slashed{p} = m_N$, one obtains \cite{Scherer:2002tk}
\eq{
	S(p) &= \frac{1}{\slashed{p}-\overset{\circ}{m}_N-\Sigma_s(p^2)-\slashed{p}F(p^2)} 
	\\
	&=
	 \Bigg[\slashed{p}-\overset{\circ}{m}_N-\left(\Sigma_s(m_N^2)+(p^2-m_N^2)\frac{\dd }{\dd p^2}\Sigma_s(p^2)\Bigr|_{p^2=m_N^2}+\cdots\right)
	\\
	&\phantom{a+a+a}-\slashed{p}\left(F(m_N^2)+(p^2-m_N^2)\frac{\dd}{\dd p^2}F(p^2)\Bigr|_{p^2=m_N^2}+\cdots\right)\Bigg]^{-1}
	\\
	&
	=
	\frac{1}{(\slashed{p}-m_N)\left[1-F(m_N^2)-2m_N^2\frac{\dd }{\dd p^2}F(p^2)\Bigr|_{p^2=m_N^2}-2m_N\frac{\dd }{\dd p^2}\Sigma_s(p^2)\Bigr|_{p^2=m_N^2}\right]}, 
} 
where the last line applies in the $\slashed{p}\rightarrow m_N$ limit. 
Thus we identify 
\eq{\label{E: Z}
	Z^{-1} = 1-F(m_N^2)-2m_N^2\frac{\dd }{\dd p^2}F(p^2)\Bigr|_{p^2=m_N^2}-2m_N\frac{\dd }{\dd p^2}\Sigma_s(p^2)\Bigr|_{p^2=m_N^2}. 
}
Modulo projections and traces, the Lorentz invariant functions $F(p^2)$ and $\Sigma_s(p^2)$ are given by (the sum of) self-energy diagrams such as the one in Figure \ref{Fig: NpiSelf_Energy}. 

Generalizing to include all the octet-baryon--meson ($OM$) and decuplet-baryon--meson ($DM$) fluctuations we write 
\eq{\label{E: ZNDelta}
	Z^{-1} = 1-\sum_{i=\{OM\},\{DM\}} X^i = 1-X, 
}
where we have defined 
\eq{\label{E: defX}
	\sum_{i=\{OM\},\{DM\}} X^i \equiv X, 
}
where the index $i$ runs over all the $OM$ and $DM$ pairs that contributes to the loop. They are collected in Table \ref{Table: Coupling}. For a given baryon-meson pair $i$, $X^i$ is given by 
\eq{\label{E: Xi}
	X^i &\equiv F^{i}(m_N^2)
	+2m_N^2\frac{\dd }{\dd p^2}F^{i}(p^2)\Bigr|_{p^2=m_N^2}
	+2m_N\frac{\dd }{\dd p^2}\Sigma_s^{i}(p^2)\Bigr|_{p^2=m_N^2}
	\\
	&
	=F^{i}(m_N^2)+m_N\frac{\dd}{\dd p_0} F^{i}(p_0^2)\Bigr|_{p_0=m_N}
	+\frac{\dd }{\dd p_0}\Sigma_s^{i}(p_0^2)\Bigr|_{p_0=m_N}, 
}
where the second line is written in the rest frame of the proton where $\vect{p} = 0$. 
We will now derive the explicit expressions for the functions $F^{i}$ and $\Sigma_s^{i}$. 

For the sake of clarity we split up the calculation in two parts, first we deal with the case where an octet-baryon and a meson is in the loop. After this we deal with the $DM$ contribution. 

The value of the self energy diagram of Figure \ref{Fig: NpiSelf_Energy} is given by  
\eq{
	-\ii\Sigma(p) = \sum_{i} \left[-\ii \Sigma^{i}(p)\right] 
	= 
	\sum_{OM} \left[-\ii \Sigma^{OM}(p)\right]
	+
	\sum_{DM} \left[-\ii \Sigma^{DM}(p)\right].  
}
Using Equation (\ref{E: scalarvector}), we can write down each term as 
\eq{\label{E: DMcase}
	\Sigma^{DM}(p) = \Sigma_s^{DM}(p^2)+\slashed{p} F^{DM}(p^2) 
}
and
\eq{\label{E: OMcase}
	\Sigma^{OM}(p) = \Sigma_s^{OM}(p^2)+\slashed{p} F^{OM}(p^2), 
}
where the scalar and vector parts can be projected out using Equation (\ref{E: projection}). 

Let us consider a generic $OM$ contribution to the self energy of the proton. The Feynman diagram for this process is shown in Figure \ref{Fig: NpiSelf_Energy} with an octet-baryon (i.e.\ $B=O$ in the figure) and a meson ($M$) in the loop. The computation is straightforward, using the Lagrangian (\ref{E: octLag}) we find 
\eq{
	-\ii \Sigma^{OM} &= -|g_{OM}|^2\int \dbar^4 k\, |\phi(\vect{k},\Lambda)|^2(\ii k_\mu )\gamma^\mu\gamma_5  \frac{\ii }{\slashed{p}-\slashed{k}-m_O+\ii\eta}
	\frac{\ii (-\ii k_{\tilde{\mu}})\gamma^{\tilde{\mu}}\gamma_5}{k^2-m_M^2+\ii\epsilon}
	\\
	&
	= +|g_{OM}|^2\int \dbar^4 k\, |\phi|^2 \frac{k_\mu k_{\tilde{\mu}}\gamma^\mu\gamma_5 
	(\slashed{p}-\slashed{k}+m_O)
	\gamma^{\tilde{\mu}}\gamma_5
	}{[k^2-m_M^2+\ii\epsilon][(p-k)^2-m_O^2+\ii\eta]},  
}
where $\dbar\equiv \dd/(2\pi)$ and $m_M$ is the mass of the meson and $m_O$ is the mass of the octet baryon. The function $\phi(\vect{k},\Lambda)$ is a form factor needed to regularize the divergencies. We will discuss $\phi$ in more detail in Section \ref{Sec: FF}. Before that let us use (\ref{E: projection}) to project out $\Sigma_s^{OM}(p^2)$ and $F^{OM}(p^2)$. We obtain 
\eq{\label{E: sigmaOM}
	\Sigma_s^{OM}(p^2) = -\ii |g_{OM}|^2 m_O\int \dbar^4 k |\phi|^2\frac{k^2}{k^2-m_M^2+\ii\epsilon}\frac{1}{(k-p)^2-m_O^2+\ii\eta}
}
and 
\eq{\label{E: fom}
	F^{OM}(p^2)= -\ii\frac{ |g_{OM}|^2}{p^2}
	\int \dbar^4k |\phi|^2\frac{k^2(k\cdot p +p^2)-2(k\cdot p)^2}{[k^2-m_M^2+\ii\epsilon][(k-p)^2-m_O^2+\ii\eta]}. 
}

The calculation for the $DM$ contributions to the self-energy loop is similar. Using the Lagrangian (\ref{E: decLag}) we find 
\eq{
	-\ii\Sigma^{DM} &= -|g_{DM}|^2\int\dbar^4 k |\phi|^2  
	\varepsilon^{\rho\mu\alpha\beta}
	\ii( p_\alpha- k_\alpha)\gamma_5\gamma_\rho 
	\ii k_\mu
	\ii G^{\text{RS}}_{\beta\tilde{\beta}}(p-k)\frac{\ii}{k^2-m_M^2+\ii\epsilon}
	\\
	&
	\phantom{= -|g_{DM}|^2\int\dbar^4 k |\phi|^2 }
	\varepsilon^{\tilde{\rho}\tilde{\mu}\tilde{\alpha}\tilde{\beta}}[-\ii(p_{\tilde{\alpha}}-k_{\tilde{\alpha}})]\gamma_5\gamma_{\tilde{\rho}}(-\ii k_{\tilde{\mu}})
	\\
	&=
	 +|g_{DM}|^2\int\dbar^4 k |\phi|^2 
	 \varepsilon^{\rho\mu\alpha\beta}
	\varepsilon^{\tilde{\rho}\tilde{\mu}\tilde{\alpha}\tilde{\beta}} 
	( p_\alpha- k_\alpha)
	(p_{\tilde{\alpha}}-k_{\tilde{\alpha}})
	k_\mu k_{\tilde{\mu}}
	\frac{\gamma_{5}\gamma_\rho G^{\text{RS}}_{\beta 
	\tilde{\beta}}
	\gamma_5\gamma_{\tilde{\rho}}}{k^2-m_M^2+\ii\epsilon}, 
	\\
}
where $G^{\text{RS}}_{\beta \tilde{\beta}}$ is the Rarita-Schwinger propagator given in Equation (\ref{E: RSPropagator}). 
From this we obtain [using the projections (\ref{E: projection})] 
\eq{\label{E: sigmadm}
	\Sigma_s^{DM}(p^2) = -\ii\frac{2 |g_{DM}|^2 m_D}{3}
	\int\dbar^4 k |\phi|^2 \frac{k^2p^2-(k\cdot p)^2}{[k^2-m_M^2+\ii\epsilon][(k-p)^2-m_D^2+\ii\eta]}
}
and 
\eq{\label{E: fdm}
	F^{DM}(p^2) = -\ii\frac{2 |g_{DM}|^2}{3p^2}
	\int\dbar^4 k |\phi|^2 \frac{(k\cdot p -p^2)((k\cdot p)^2-k^2 p^2)}{[k^2-m_M^2+\ii\epsilon][(k-p)^2-m_D^2+\ii\eta]} 
}
where $m_D$ is the mass of the decuplet baryon. 
These expressions can further be simplified by adding zero to the numerator so as to cancel one of the propagators, with the cost of having some extra one-propagator terms. This amounts to the rewriting $k^2 = \Delta_M+m_M^2$ and $k\cdot p =\frac{1}{2}\Delta_M-\frac{1}{2}\Delta_B+\frac{1}{2}\left(p^2+m_M^2-m_B^2\right)$ and using the fact that any odd term in $k$ integrates to zero, see Appendix \ref{Sec: arcInfinity}. Here $\Delta_M$ and $\Delta_B$ denote the denominators of the meson and baryon propagators respectively. 

We will work in the rest frame of the proton, $\vect{p} = 0$, and do the $k_0$-integrals with residue techniques. Thus what we need in the end for Equation (\ref{E: Xi}) are the derivatives of the functions $\Sigma_s^{OM}(p_0^2)$, $F^{OM}(p_0^2)$, $\Sigma_s^{DM}(p_0^2)$ and $F^{DM}(p_0^2)$ with respect to $p_0$ evaluated at $p_0=m_N$. 
\begin{figure}[t]
	\includegraphics[width=0.7\linewidth]{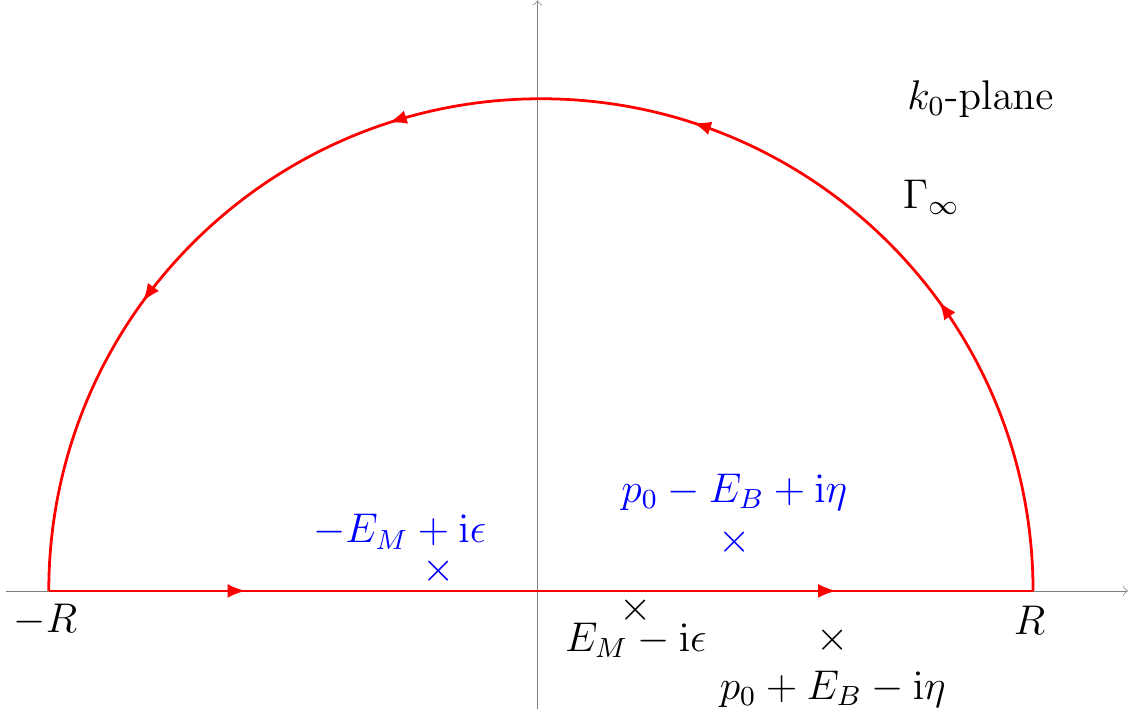}
	\caption{The poles of the functions $F^{BM}(p_0^2)$ and $\Sigma_s^{BM}(p_0^2)$ and the contour of integration in the complex $k_0$-plane. }\label{Fig: Poles}
\end{figure}

The poles of the meson and baryon propagators, i.e.\ the zeros of $\Delta_M$ and $\Delta_B$ respectively, are located at 
\eq{
	k_0^2-E_M^2+\ii\epsilon &= 0 \leadsto [k_0-E_M+\ii\epsilon][k_0+E_M-\ii\epsilon]\approx 0
	\\
	&\Rightarrow \begin{cases}
k_0= -E_M+\ii\epsilon,~\text{(located in the upper half-plane)},\\
k_0=+E_M-\ii\epsilon,~\text{(located in the lower half-plane)} 
\end{cases}
}
and at 
\eq{
	(k_0-p_0)^2-E_B^2+\ii\eta &= 0 \leadsto [k_0-p_0-E_B+\ii\eta][k_0-p_0+E_B-\ii\eta]\approx 0
	\\
	&\Rightarrow \begin{cases}
k_0=p_0-E_B+\ii\eta,~\text{(located in the upper half-plane)},\\
k_0=p_0+E_B-\ii\eta,~\text{(located in the lower half-plane)}.  
\end{cases}
}
Here we have defined the meson and baryon energies by $E_M  = \sqrt{\vect{k}^2+m_M^2}$ and $E_B  = \sqrt{\vect{k}^2+m_B^2}$ respectively. The poles and the contour of integration are shown in Figure \ref{Fig: Poles}. Then, Equation (\ref{E: sigmaOM}) can be written [recalling $\dbar=\dd/(2\pi)$]  
\eq{
	\Sigma_s^{OM}(p^2) = \frac{|g_{OM}|^2 m_O}
	{\left(2\pi\right)^3}\int \dd^3 k |\phi|^2 \sum_{\text{res}}
	\Bigg[
	\frac{k_0^2-\vect{k}^2}{[k_0^2-E_M^2+\ii\epsilon][(k_0-p_0)^2-E_O^2+\ii\eta]};j\Bigg], 
}
where we pick up the residues at the poles $j$, marked blue in Figure \ref{Fig: Poles}, located in the upper half-plane enclosed by the contour $\Gamma = (-R,R)+\Gamma_\infty$ with $R\rightarrow \infty$. Applying the same technique to (\ref{E: fom}), (\ref{E: sigmadm}) and (\ref{E: fdm}) we finally obtain (in the proton rest frame) 
\eq{\label{E: XiExplicit}
	X^{i}(\Lambda) =   
	\int_{0}^{\infty} \frac{\dd k k^2 \left|\phi\left(k,\Lambda\right)
	\right|^2}{2\pi^2}
	|g_i|^2 \xi_i(k;\, m_B,m_M,m_N)
}
where $k$ denotes the modulus of the 3-momentum:\ $k\equiv \lVert \vect{k} \rVert$ and we have taken $\phi$ to be rotationally invariant i.e.\ $\phi({\vect{k}}) = \phi(k)$ and used 
\eq{\label{E: Meas}
	\int\dbar^3 k \equiv \int \frac{\dd^3k}{(2\pi)^3} = \int\frac{{k}^2\,\dd k \,\dd\Omega}{(2\pi)^3} = \frac{4\pi}{8\pi^3}\int_0^\infty\dd k\,{k}^2 = \frac{1}{2\pi^2}\int_0^\infty \dd k\, {k}^2. 
} 
For the octet-baryon--meson part, the functions $\xi_i(k)$ are given by 
\eq{\label{E: xiOM}
		\xi_{OM} = 
	\frac{2m_O m_N m_M^2(E_O+E_M)-\left[2k^2(E_O+E_M)+m_M^2E_O\right]\left[(E_O+E_M)^2+m_N^2\right]}{2E_OE_M[m_N^2-(E_O+E_M)^2]^2}, 
}
where $OM\in \{n\pi^+,P\pi^0,\Lambda K^+, \Sigma^+ K^0, \Sigma^0 K^+, P\eta \}$. 
For the decuplet contribution, the functions $\xi_i(k)$ are given by 
\eq{\label{E: xiDM}
	\xi_{DM} = m_N k^2\, \frac{ -2 m_D
   (E_D+E_M)^3-3 E_D m_N (E_D+E_M)^2+E_D
   m_N^3}{3 E_D E_M [m_N^2-(E_D+E_M)^2]^2
  }
}
with $DM \in \{\Delta^{++}\pi^-,\Delta^{+}\pi^0,\Delta^{0}\pi^+,\Sigma^{*+}K^0,\Sigma^{*0}K^+\}$. It should be clear that $\xi_{OM}$ and $\xi_{DM}$ are functions of $k$ and the relevant masses. Some comments on $X^i$ and its integrand are in order. 

Let us recall that $Z$ is the probability to find the bare proton in the physical proton wave function, cf.\ Equation (\ref{E: FockExp}). Therefore the probability to find all the fluctuations we have taken into account is given by $1-Z$. Using Equation (\ref{E: ZNDelta}) and (\ref{E: defX}) we can separate it into its constituent parts 
\eq{
	1-Z = 1- \frac{1}{1-X} = \frac{-\sum_i X^i}{1-X} = \frac{-X^{n\pi^+}}{1-X}+\frac{-X^{P\pi^0}}{1-X} + \cdots, 
}
where the ellipsis include all the other $OM$ and $DM$ pairs listed in Table \ref{Table: Coupling}. Thus, each term $-X^i/(1-X)$ gives the probability to find that specific fluctuation in the wave function of the physical proton.\footnote{Notice that from the definition of $Z$ Equation (\ref{E: ZNDelta}) follows that each $X^i<0$, which is indeed the case here as can be checked. } Or in other words we have now the explicit expressions for the coefficients in Equation (\ref{E: FockExp})
\eq{\label{E: alpha2}
	|\alpha_{n\pi^+}|^2 = \frac{-X^{n\pi^+}}{1-X}, 
	~~~~~~|\alpha_{P\pi^0}|^2 = \frac{-X^{P\pi^0}}{1-X}, ~~~~~~\text{and so on. }
} 

Now let us consider the functions $\xi_i(k, m_B, m_M, m_N)$ that appear in the integrand of $X^i$. Using their explicit form given by equations (\ref{E: xiOM}) and (\ref{E: xiDM}), one can write down their large-$k$ behavior 
\eq{
	\xi_{OM} \rightarrow -\frac{1}{2k}, ~~~~~\text{(as $k\rightarrow \infty$)}
}
and 
\eq{
	\xi_{DM}\rightarrow -\frac{m_N (4 m_D+3 m_N)}{12 k}, ~~~~~\text{(as $k\rightarrow \infty$)}. 
}
The difference in dimensionality is due to the fact that the mass dimensions of the couplings are different, namely $[g_{OM}] = -1$ while $[g_{DM}] = -2$. The point is that the large-$k$ behavior $1/k$ exhibited by both $\xi_{OM}$ and $\xi_{DM}$ is not sufficient to make the integrals, $X^i$, convergent.  Therefore one needs to regularize the integrals of Equation (\ref{E: XiExplicit}) and we do this by introducing a form factor which is the topic of the next section. 
\section{Form Factor}\label{Sec: FF}
As with any effective theory the hadronic description of the microscopic world has a limited range of applicability. This range in momentum space does not have a sharp boundary, but a good rule of thumb is that momentum transfers larger than 1 GeV are outside of the applicability range. Thus from a physical point of view, it is natural to cut off the high-momentum part of the hadronic fluctuations. The effective nature of the theory (the Lagrangians) manifests itself in the divergency of various quantities. To regularize these we choose to introduce a form factor at each vertex that smoothly cuts off the momenta involved. 

As shown in the previous sections, the $k_0$-integration can be safely done using residue techniques hence only the $\dd^3k$-integral needs to be regularized here. One can then choose to cut off the 3-momentum of one of the hadrons in the fluctuation as seen in the rest frame of the proton. However, in other applications of the hadronic language, such as e.g.\ in convolution models in deep inelastic scattering (DIS) on the proton, the form factors used must satisfy certain conditions in order for quark number, momentum and other sum-rules to work out properly. It turns out \cite{AESLGIHGasymm,AESLGIHGspin} that an admissible form factor depending on the combination ${k}_1^2+{k}_2^2$, where ${k}_{1,2}^2$ are the squares of the 3-momenta of the particles in the fluctuation, is sufficient to not ruin any of the sum-rules. A natural choice is then a form factor depending on the average of the squares of the baryon and meson 3-momentum, as seen in the rest frame of the proton. 

One can choose this form factor in many ways but one simple ansatz is to use a standard Gaussian. As we will see in the following section using a Gaussian form factor is natural in the sense that it yields a near one-to-one relation between the cut-off parameter $\Lambda$ and the average momentum distribution in the fluctuation, which makes interpreting the value of $\Lambda$ more straightforward. 

For the case at hand ${k}_{1}^2 ={k}_{2}^2 = {k}^2$ resulting in the form factor 
\eq{\label{E: FF1}
	\phi_1(k,\Lambda) = \exp\left[-\frac{k^2}{\Lambda^2}\right], 
} 
which we will use. 
\begin{figure}
	 \includegraphics[width = 1\linewidth]{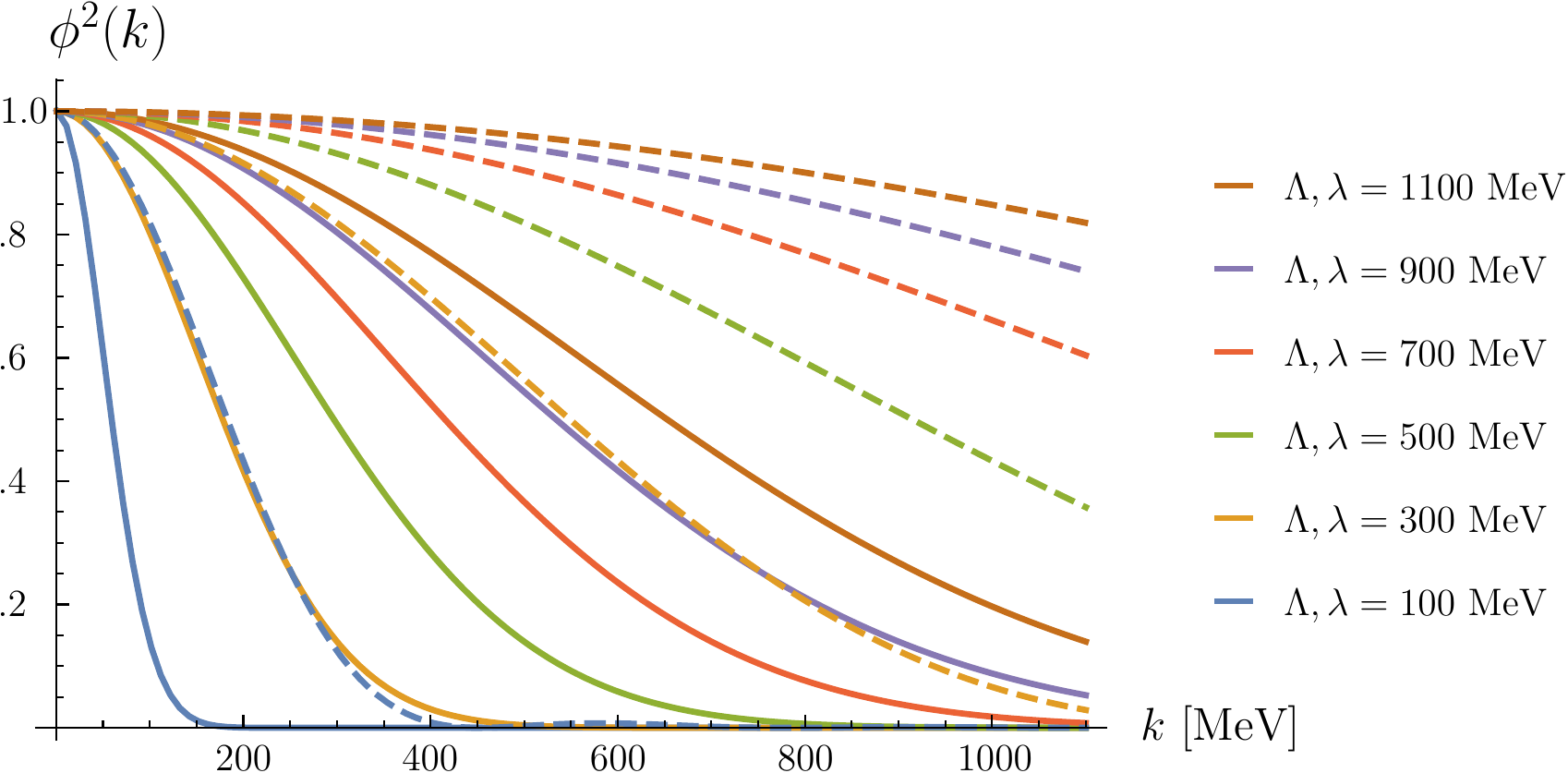}\caption{The square of each form factor as a function of $k$ and for different values of $\Lambda,\lambda$. Solid (dashed) curves are $\phi_1^2$ ($\phi_2^2$). }\label{Fig: FormFactors}
\end{figure}

For sake of comparison we will also include a more extreme form factor, namely one corresponding to a hard boundary in position space, i.e.\ $\tilde{\phi_2}(\vect{x}) \propto 1/R^3$ for $||\vect{x}||\leq R$ and zero elsewhere simulating a sphere of radius $R$ inside of which the fluctuating particles are confined.\footnote{The reason for the form of the definition of $\tilde{\phi_2}(\vect{x})$ is because we want its Fourier transform to be dimensionless. } Since we are working in momentum space we Fourier transform $\tilde{\phi_2}(\vect{x})$ and find it to be proportional to a Bessel function $\,\phi_2\propto(k/\lambda)^{-3/2}J_{3/2}(k/\lambda)$ which we normalize such that it yields unity for zero fluctuation momentum: 
\eq{\label{E: FF2}
	\phi_2(k,\lambda) = \mathcal{N}\frac{\sin\left(\frac{k}{\lambda}\right)-\left(\frac{k}{\lambda}\right)\cos\left(\frac{k}{\lambda}\right)}{\left(\frac{k}{\lambda}\right)^3}, 
}
with $\mathcal{N} = 3$ and we have taken 
	$R=1/\lambda$.  
Both of these form factors are shown in Figure \ref{Fig: FormFactors} where we have plotted their squares for various values of $\Lambda$ and $\lambda$. We will comment more on the effects of these form factors in Section \ref{Sec: Results}. 


\section{Results}\label{Sec: Results}
Having derived the relevant formulas we now present the results. For the couplings we use the following values 
	$D = 0.8, 
	F = 0.46, 
	F_\pi = 92.4 ~\text{MeV}$ and $h_A = 2.67\pm 10\%$, see Equation (\ref{E: hupperlower}) and below it for more details and references. 
\begin{figure*}[!t]
	\subfloat[]{\label{Fig: 
	indprob1a}
	\includegraphics[width = 0.99\linewidth]{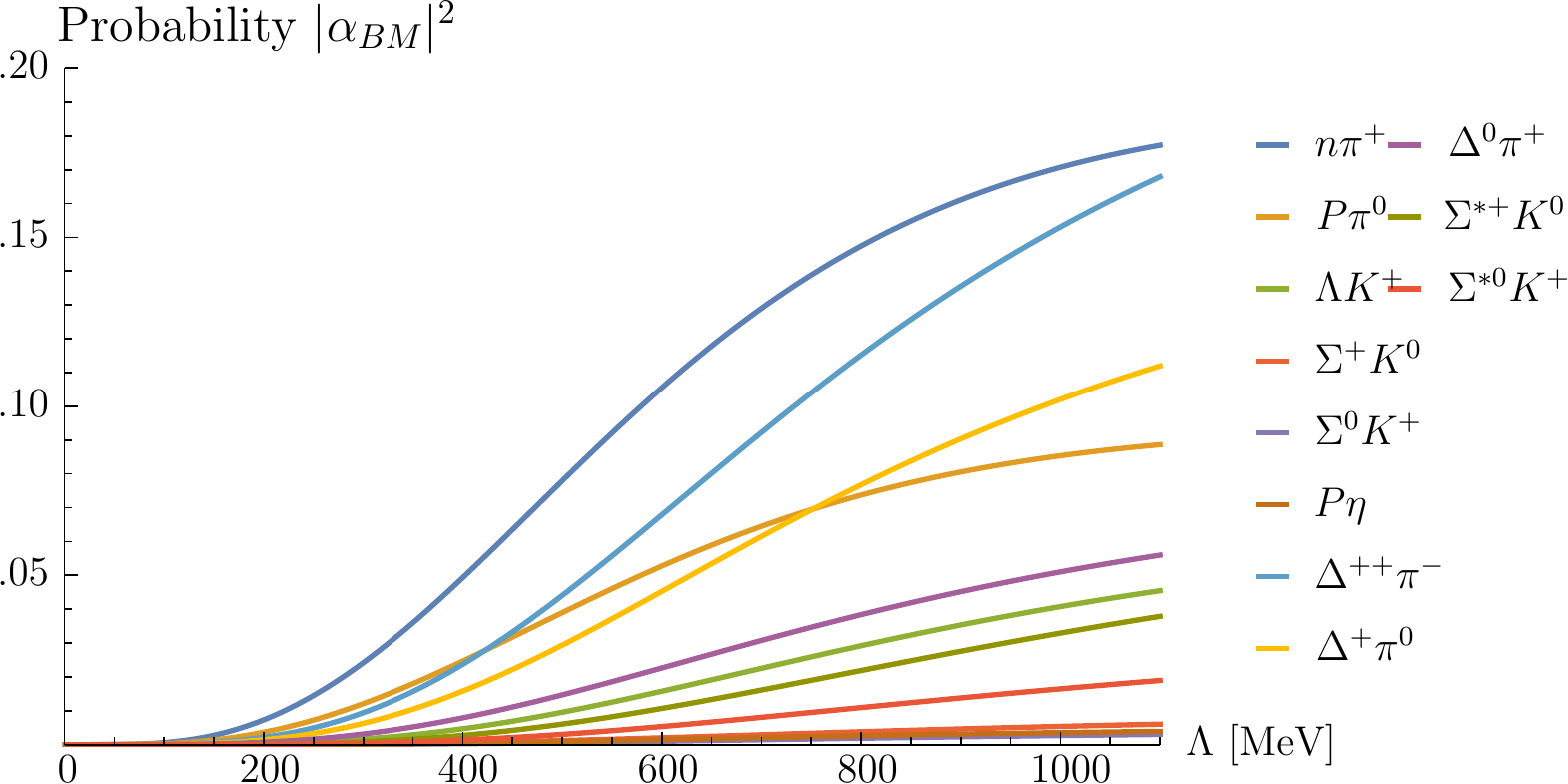}	
	}
	~
	\\
	\subfloat[]{\label{Fig: indprob2a}
		\includegraphics[width = 0.99\linewidth]{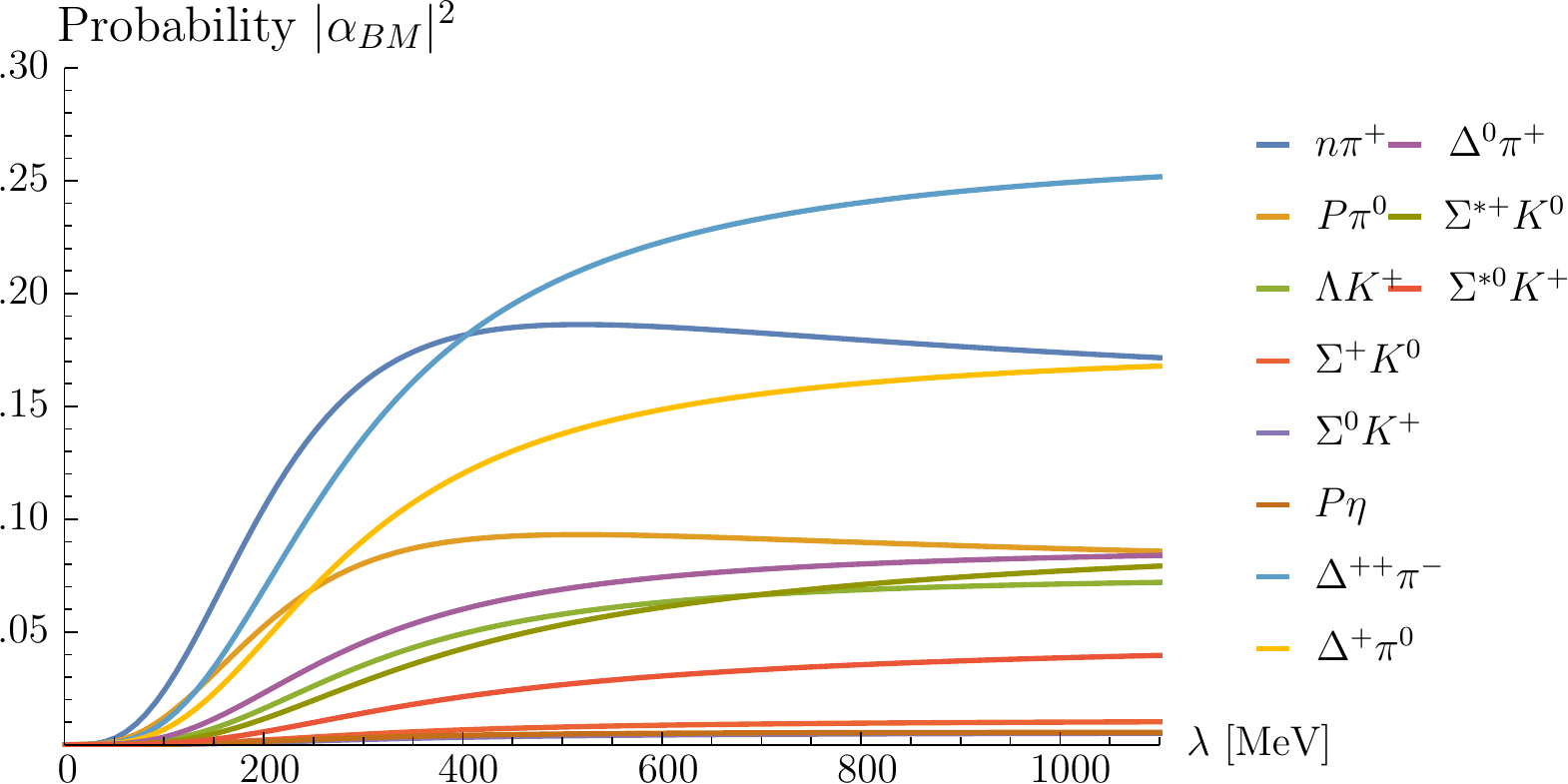}
	}
	\caption{The different fluctuation 
	probabilities $|\alpha_{BM}|^2$ as a function of the cut-off parameter. In \protect\subref{Fig: indprob1a} using the form factor $\phi_1(k,\Lambda) = \exp\left[-k^2/\Lambda^2\right]$, in \protect\subref{Fig: indprob2a} using the form factor $\phi_2(k,\lambda) = 3\left[\sin\left(\frac{k}{\lambda}\right)-\left(\frac{k}{\lambda}\right)\cos\left(\frac{k}{\lambda}\right)\right]/\left(\frac{k}{\lambda}\right)^3$.}\label{Fig: indprob}
\end{figure*}
The values for the different hadron masses are tabulated in Table \ref{Table: masses}. 

Recall from Equation (\ref{E: alpha2}) that the probability to find a specific baryon-meson pair, $BM$, in the proton wave function is given by $|\alpha_{BM}|^2 = -X^{BM}/(1-X)$ where $X=\sum_i X^i$ and the functions $X^i$ depend on the cut-off [c.f.\ Equation (\ref{E: XiExplicit})]. Thus the probabilities 
$|\alpha_{BM}|^2$ are also functions of the cut-off. This dependence is shown in Figure \ref{Fig: indprob1a} and \ref{Fig: indprob2a} using the form factors (\ref{E: FF1}) and (\ref{E: FF2}) respectively. Notice that since we assume perfect isospin symmetry, i.e.\ $m_{\Delta^{++}}=m_{\Delta^{+}}=m_{\Delta^{0}}$ (c.f.\ Table \ref{Table: masses}), the ratio between two probabilities from the same isospin multiplet is a constant and it is given by the ratio of the coupling constants. For instance 
\eq{
	\frac{|\alpha_{\Delta^{0}\pi^{+}}|^2}{|\alpha_{\Delta^{++}\pi^{-}}|^2} = \frac{|g_{\Delta^{0}\pi^{+}}|^2}{|g_{\Delta^{++}\pi^{-}}|^2} = \frac{1}{3}, 
}
and so on for the other isospin partners. This relation is also seen in the plots in Figure \ref{Fig: indprob1a} and \ref{Fig: indprob2a}. What is non-trivial is e.g.\ the relation between the probabilities of $\Delta\pi$ and $N\pi$. There is no simple estimate of this using just the couplings and/or mass suppression. These probabilities are interesting from the point of view of explaining the properties of the proton such as e.g.\ the $\bar{d}$-$\bar{u}$ asymmetry observed in the quark momentum distribution of the proton sea \cite{Towell:2001nh}. 

\begin{figure}[h!]
    \centering
    \subfloat[]{\label{Fig: eachContr1}
	\includegraphics[width = 0.93\linewidth]{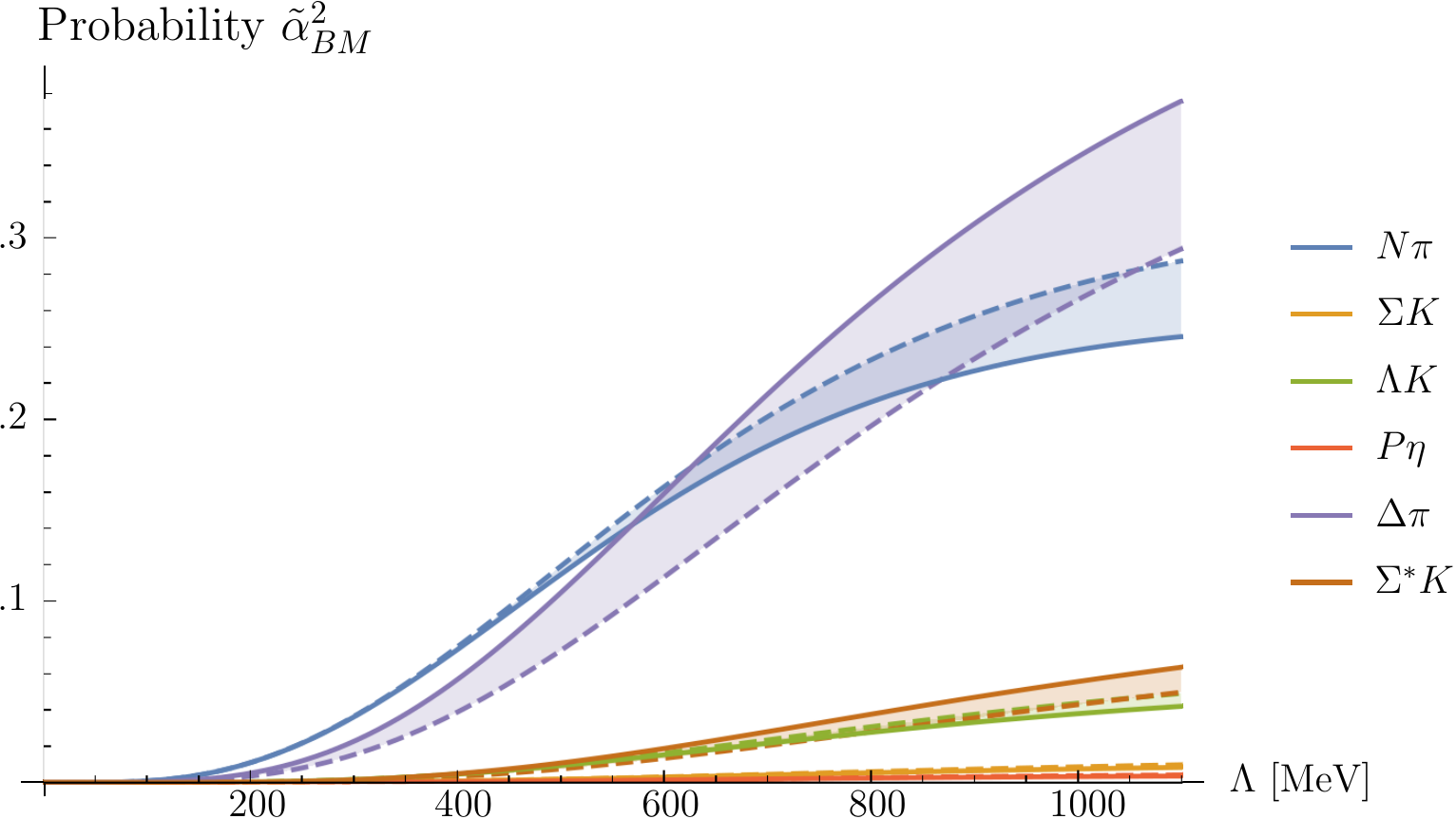}
	}
	\\
	\subfloat[]{\label{Fig: eachContr2}
	\includegraphics[width = 0.93\linewidth]{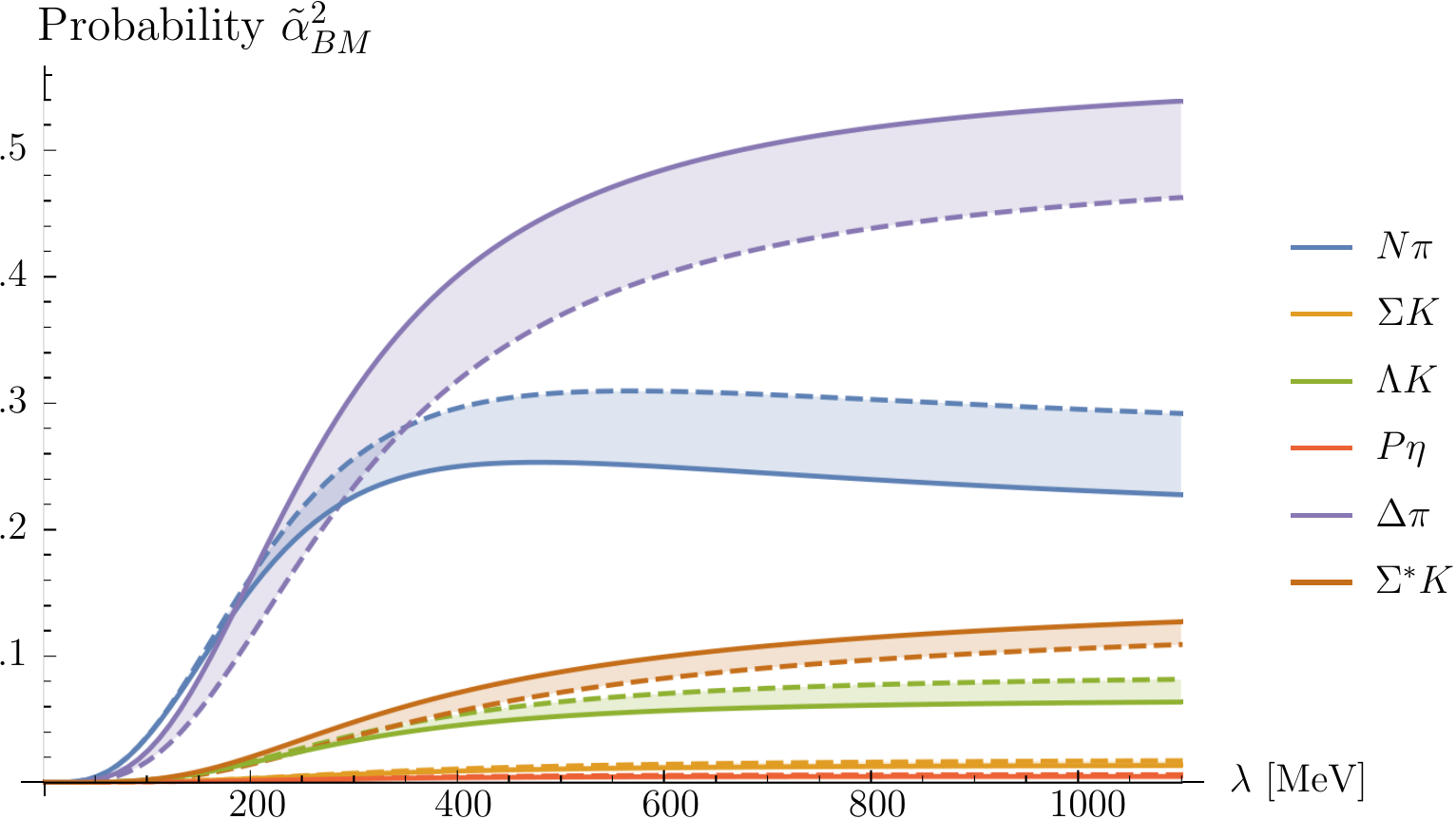}
	}
	\caption{The isospin-summed fluctuation probabilities $\tilde{\alpha}_{BM}^2$ as a function of the cut-off using $\phi_1$ in (a) and $\phi_2$ in (b). The bands are due to a variation in $h_A$: The solid (dashed) curves are for the larger $h_A^\text{max} = 2.937$ (smaller $h_A^\text{min} = 2.403$) value of $h_A$. }
\end{figure}

It is perhaps more natural to consider the isospin-summed probabilities defined as the sum of probabilities for the particles belonging to the same isospin multiplet. Thus for instance the $P\pi^0$ and the $n\pi^+$ fluctuation probabilities would be considered as the the nucleon-pion ($N\pi$) probability $\tilde{\alpha}_{N\pi}^2$ defined as 
\eq{
	\tilde{\alpha}_{N\pi}^2 \equiv |\alpha_{n\pi^+}|^2+|\alpha_{P\pi^0}|^2 = \frac{-X^{n\pi^+}-X^{P\pi^0}}{1-X}. 
}
The same goes for the other isospin partners such as e.g.\ $\tilde{\alpha}_{\Delta\pi}^2 = |{\alpha}_{\Delta^{++}\pi^{-}}|^2+|{\alpha}_{\Delta^{+}\pi^0}|^2+|{\alpha}_{\Delta^{0}\pi^+}|^2$. 
These isospin-summed probabilities are shown in Figure \ref{Fig: eachContr1} and \ref{Fig: eachContr2} using the form factors $\phi_1$ and $\phi_2$ respectively. As can be seen from these figures, the hadron cloud consists mostly of $N\pi$ and $\Delta\pi$ throughout the whole range of each respective cut-off parameter. 
What is also shown in the figures just mentioned is the effect a $10\%$ variation of the decuplet coupling $h_A$ has on the results (the shaded band). We have included this due to the uncertainty in the value of $h_A$ obtained when matched to experimental data [see the discussion above Equation (\ref{E: hupperlower})]. 
The solid (dashed) curves are those corresponding to using the larger (smaller) value for $h_A$ given by $h_A^\text{max} = 2.937$ ($h_A^\text{min} = 2.403$). As seen a $10\%$ variation in $h_A$ yields a similar variation in the $\Delta\pi$ probability and a somewhat smaller change for the $N\pi$ one. In particular when using the Gaussian form factor (cf.\ Figure \ref{Fig: eachContr1}). 
The $N\pi$ term is only indirectly affected due to the total fluctuation probability
\eq{\label{E: totFlucProb}
	1-Z =  \tilde{\alpha}_{\Delta\pi}^2
	+ \tilde{\alpha}_{\Sigma^{*}K}^2
	+ \tilde{\alpha}_{N\pi}^2
	+ \tilde{\alpha}_{\Lambda K}^2
	+ \tilde{\alpha}_{\Sigma K}^2
	+ \tilde{\alpha}_{P\eta}^2
}
being bounded from above by 1.  This is more easily seen in the larger $\lambda$ regions of Figure \ref{Fig: eachContr2} where $\tilde{\alpha}_{N\pi}^2$ actually decreases due to an increase of the other (decuplet) `channels'. 

\begin{figure}
	 \includegraphics[width = 1\linewidth]{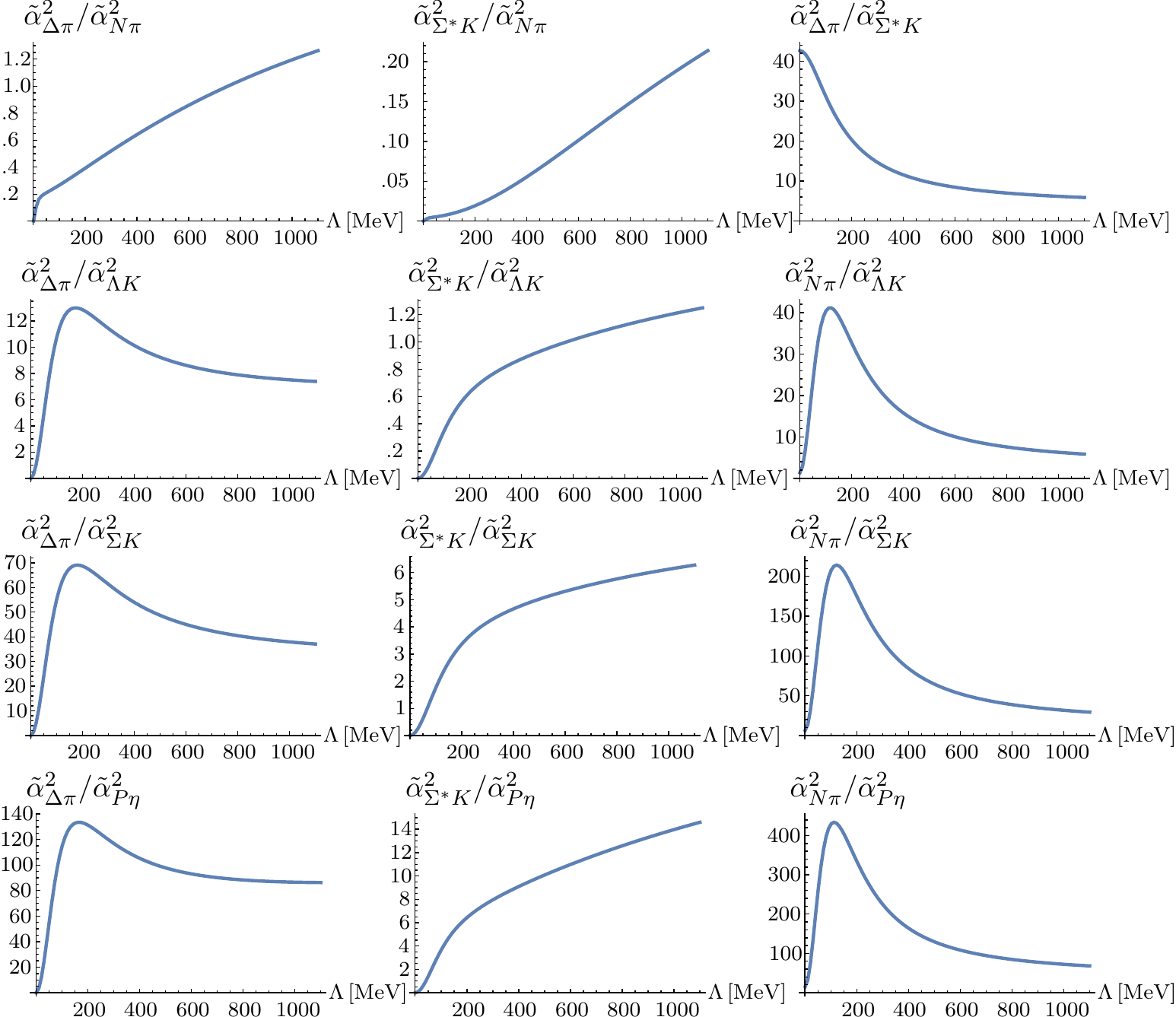}\caption{The ratios of the isospin-summed probabilities as a function of $\Lambda$. }\label{Fig: probratios}
\end{figure}

As seen in these figures there are some non-trivial correlations between the different probabilities. It is therefore interesting to study the ratios of the isospin-summed probabilities given by 
\eq{\label{E: probratios}
	\frac{\tilde{\alpha}_{BM}^2}{\tilde{\alpha}_{B'M'}^2}
}
where the mesons $M$ and $M'$ need not necessarily be different. The functions defined by (\ref{E: probratios}) are shown in Figure \ref{Fig: probratios} plotted as a function of $\Lambda$ using the Gaussian form factor $\phi_1$. 

The two form factors $\phi_1$ and $\phi_2$ have indeed different properties, in particular the latter one being oscillatory in nature. But these oscillations are rather small in amplitude relative to $\max(\phi_2) = 1$. Hence they are even smaller for the squared form factor $\phi_2^2$ that appears in the formulas. This can be seen from Figure \ref{Fig: FormFactors} where we see that $\phi_2^2(k)|_{\lambda=100 ~\text{MeV}}$ nearly overlaps with $\phi_1^2(k)|_{\Lambda=300 ~\text{MeV}}$. Similarly for the higher cut-off values e.g.\ $\phi_2^2(k)|_{\lambda=300~\text{MeV}}\approx \phi_1^2(k)|_{\Lambda=900~\text{MeV}}$. Thus naively one would think that the form factors are as good as being equivalent. 
But recall that $\phi_2$ imposes a sharp cut-off in position space which means  a lot of momentum `leakage' into the region supposedly cut off by $\phi_2(k)$ in momentum space. In other words from the point of view of an effective theory, the Besselian form factor $\phi_2(k)$ gives too much weight to the higher $k$-values. 
To show this we first study the momentum distribution given by the integrand of Equation (\ref{E: defX}). 
That is, we define $\chi(k,\Lambda)\equiv \sum_i \chi_i(k,\Lambda)$ by 
\eq{
	X = \int_0^\infty\! \dd k\, \chi(k,\Lambda), 
} 
where 
\eq{\label{E: chi}
	\chi_i(k,\Lambda) = |g_i|^2\frac{ k^2 \left|\phi\left(k,\Lambda\right)
	\right|^2}{2\pi^2}
	 \xi_i(k;\, m_B,m_M,m_N). 
}

\begin{figure}[t!]
	\subfloat[]{\label{Fig: kDist1}
	 \includegraphics[width = 0.99\linewidth]{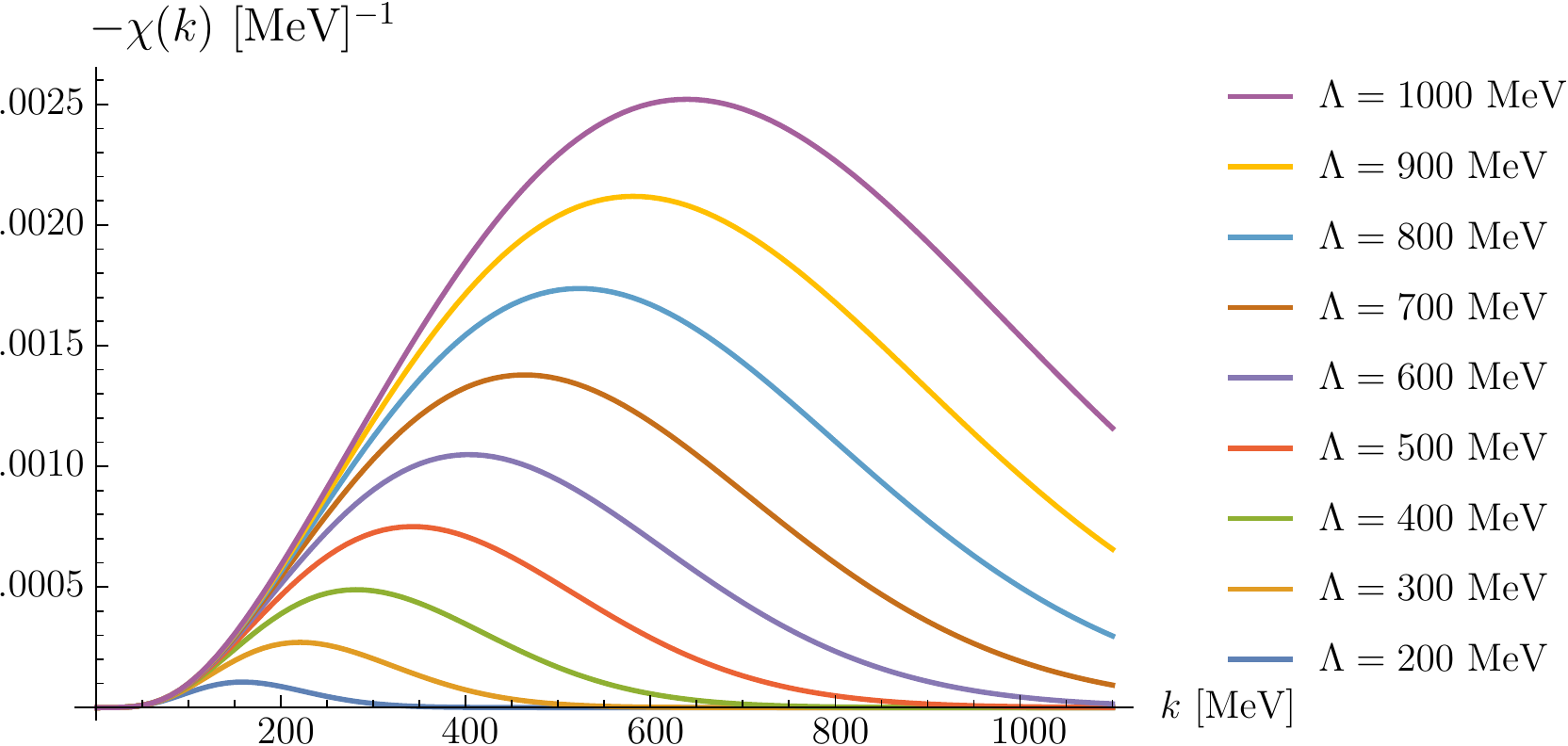}	 
	 }
	 ~
	 \\
	 \subfloat[]{\label{Fig: kDist2}
	 \includegraphics[width = 0.99\linewidth]{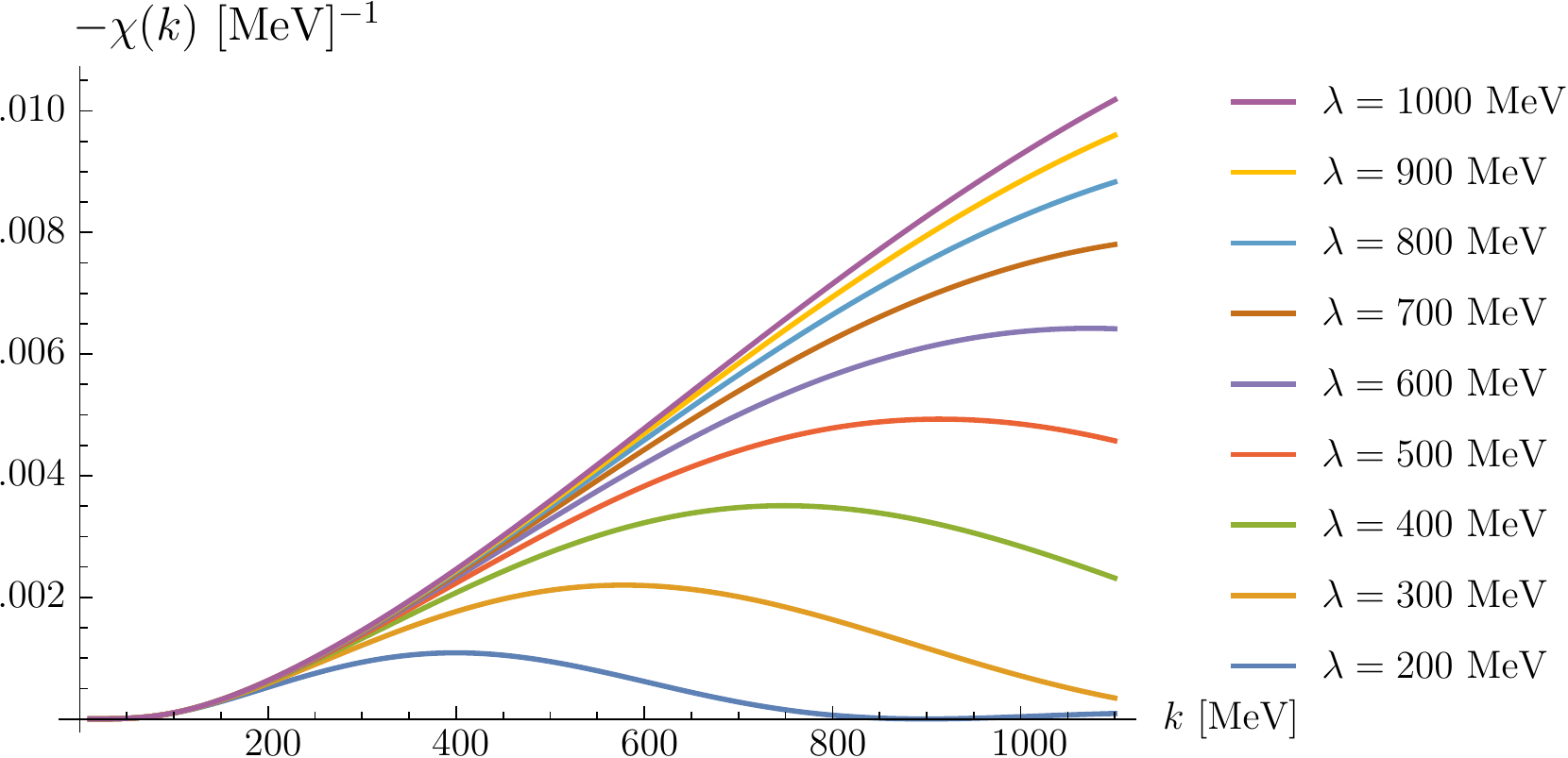}
	 }
	 \caption{The momentum distribution in the fluctuations using the form factor $\phi_1$ and $\phi_2$ in (a) and (b) respectively, for various values of the cut-off parameter. }\label{Fig: kDist}
\end{figure}

Notice that $1/(1-X)$ being a probability it must satisfy $0\leq1/(1-X)\leq1$, which it does, hence this implies that $X\leq0$. Therefore $\chi(k,\Lambda)\leq0$ which can be seen from an inspection of the explicit forms of Equation (\ref{E: chi}), (\ref{E: xiOM}) and (\ref{E: xiDM}). 

For the convenience of the reader we have plotted the negative of $\chi(k)$. This is shown in Figure \ref{Fig: kDist1} and \ref{Fig: kDist2} for various values of the cut-off parameters using the form factor $\phi_1$ and $\phi_2$ respectively. The relations between the form factors e.g.\ $\phi_2^2(k)|_{\lambda=300~\text{MeV}}\approx \phi_1^2(k)|_{\Lambda=900~\text{MeV}}$ is reflected in the peak position in the momentum distributions they yield. For instance in Figure \ref{Fig: kDist1} we see that the momentum distribution peaks at $k=600~\text{MeV}$ for $\Lambda=900~\text{MeV}$, while the same holds true at $\lambda = 300 ~\text{MeV}$ using the form factor $\phi_2$, see Figure \ref{Fig: kDist2}. 

What really is appealing with the Gaussian form factor though is that it yields a straightforward interpretation of the cut-off parameter as the average momentum distribution in the fluctuation. This can be seen by considering the average momentum distribution $\kappa = \sqrt{\langle k^2\rangle }$ defined by 
\eq{\label{E: kappa}
	\kappa^2(\Lambda) = \frac{\int_0^\infty\! \dd k\, k^2  \chi(k,\Lambda)}{\int_0^\infty\! \dd k\,  \chi(k,\Lambda)}. 
} 
\begin{figure*}[!t]
    \centering
    \subfloat[]{\label{Fig: kappa1}
        \centering
        \includegraphics[width = 0.48\linewidth]{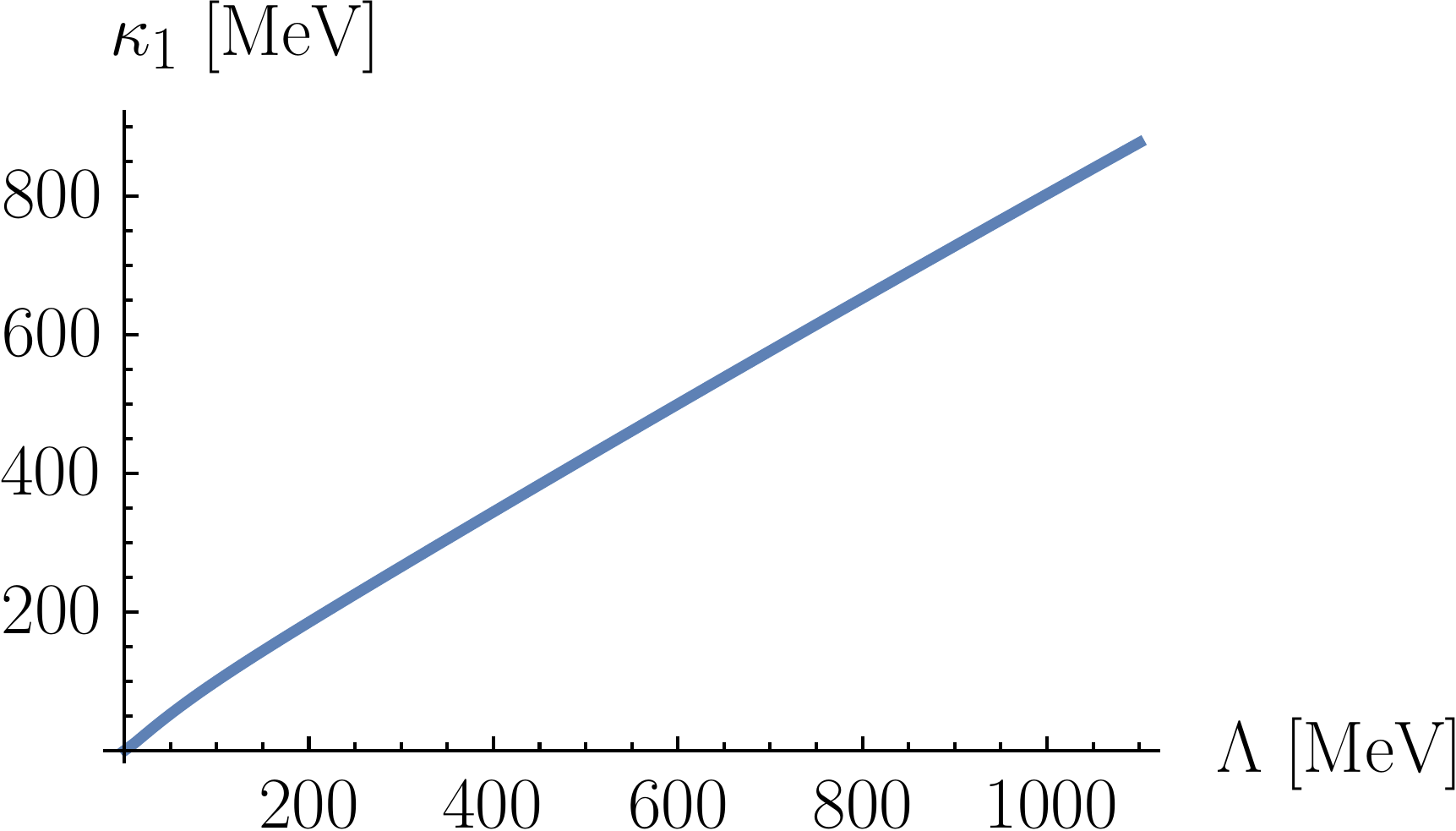}
    }%
    ~ 
    \subfloat[]{\label{Fig: kappa2}
        \centering
        \includegraphics[width = 0.48\linewidth]{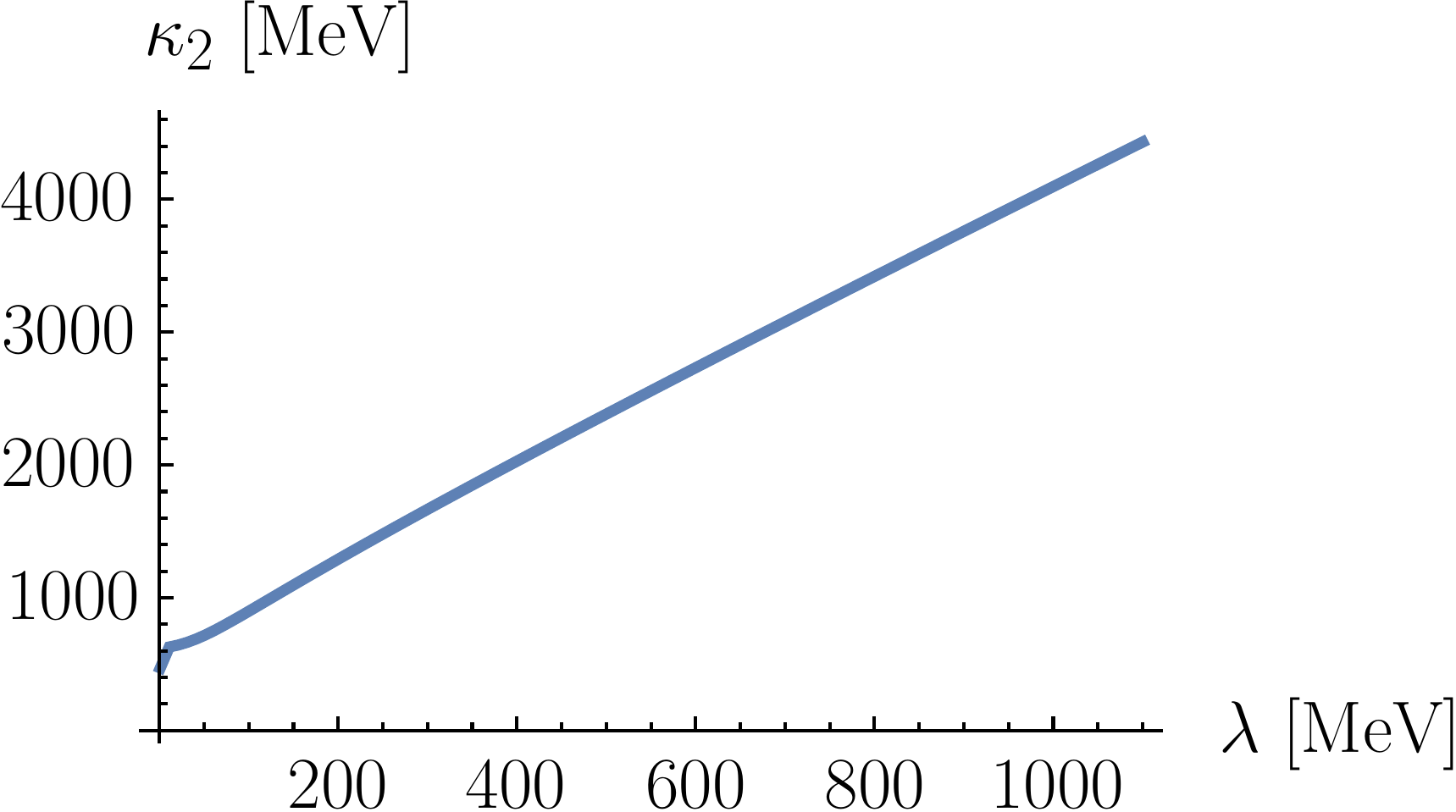}
    }
    \caption{The average momentum distribution $\kappa$ defined in Equation (\ref{E: kappa}) as a function of the cut-off parameter. }\label{Fig: kappa} 
\end{figure*}
In figures \ref{Fig: kappa1} and \ref{Fig: kappa2} we show $\kappa_1(\Lambda)$ and $\kappa_2(\lambda)$ when using the form factor $\phi_1$  and $\phi_2$ respectively. As can be seen, when using the Gaussian form factor $\phi_1$, the resulting average momentum distribution satisfies $\kappa_1(\Lambda)\approx \Lambda$, i.e.\ the slope is near unity. This is not the case when using the Besselian form factor $\phi_2$ which as seen in Figure \ref{Fig: kappa2} results in a much steeper slope (notice the different scales on the vertical axis). This reflects the oscillatory nature of the Besselian form factor which puts more weight on larger $k$-values than does the Gaussian. Thus, from an effective theory point of view, the Gaussian form factor is more attractive. 

\begin{figure*}[!t]
    \centering
    \subfloat[]{\label{Fig: ratioLowUp1Plot}
        \centering
        \includegraphics[width = 0.48\linewidth]{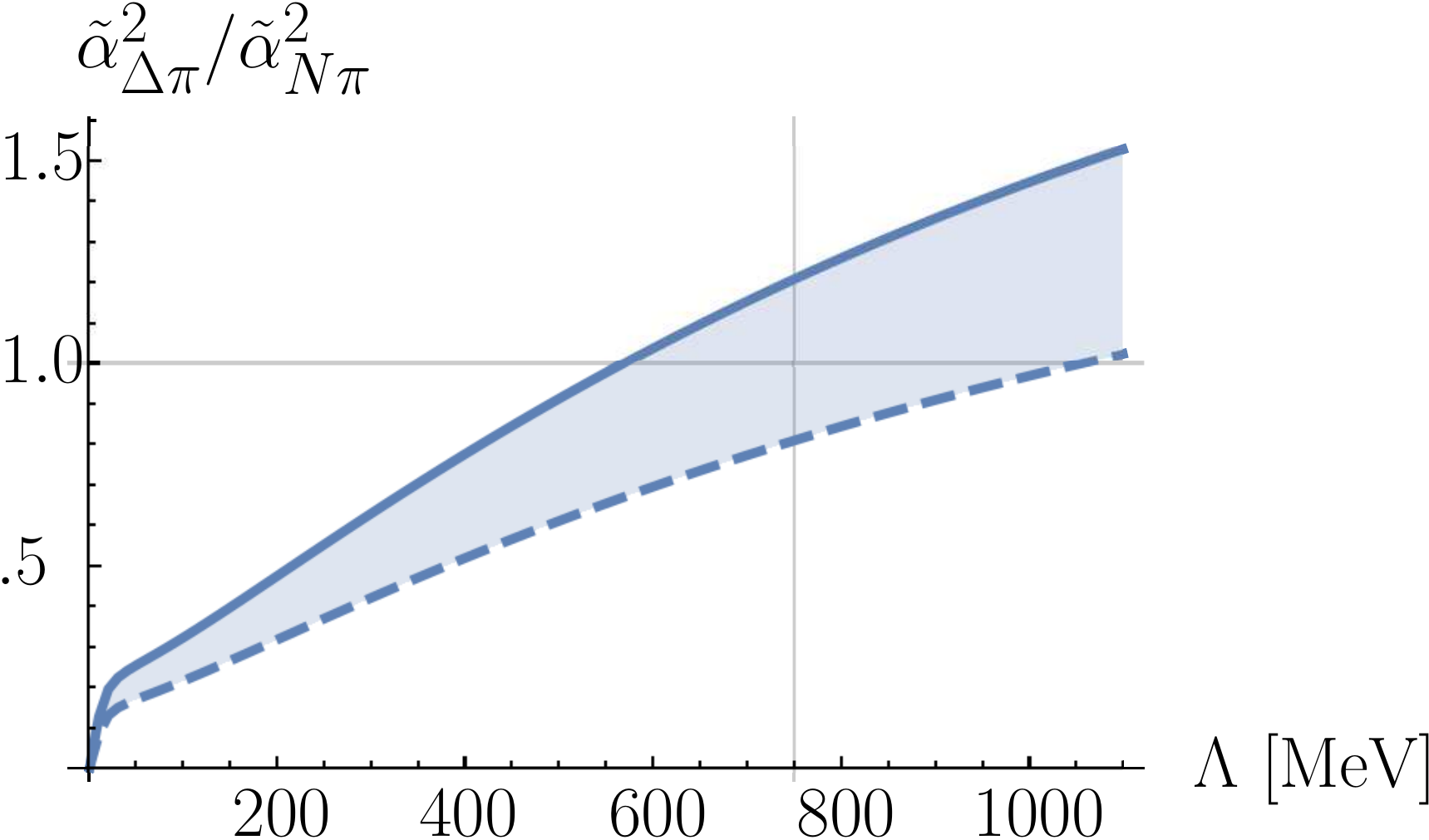}
    }%
    ~ 
    \subfloat[]{\label{Fig: ratioLowUp2Plot}
        \centering
        \includegraphics[width = 0.48\linewidth]{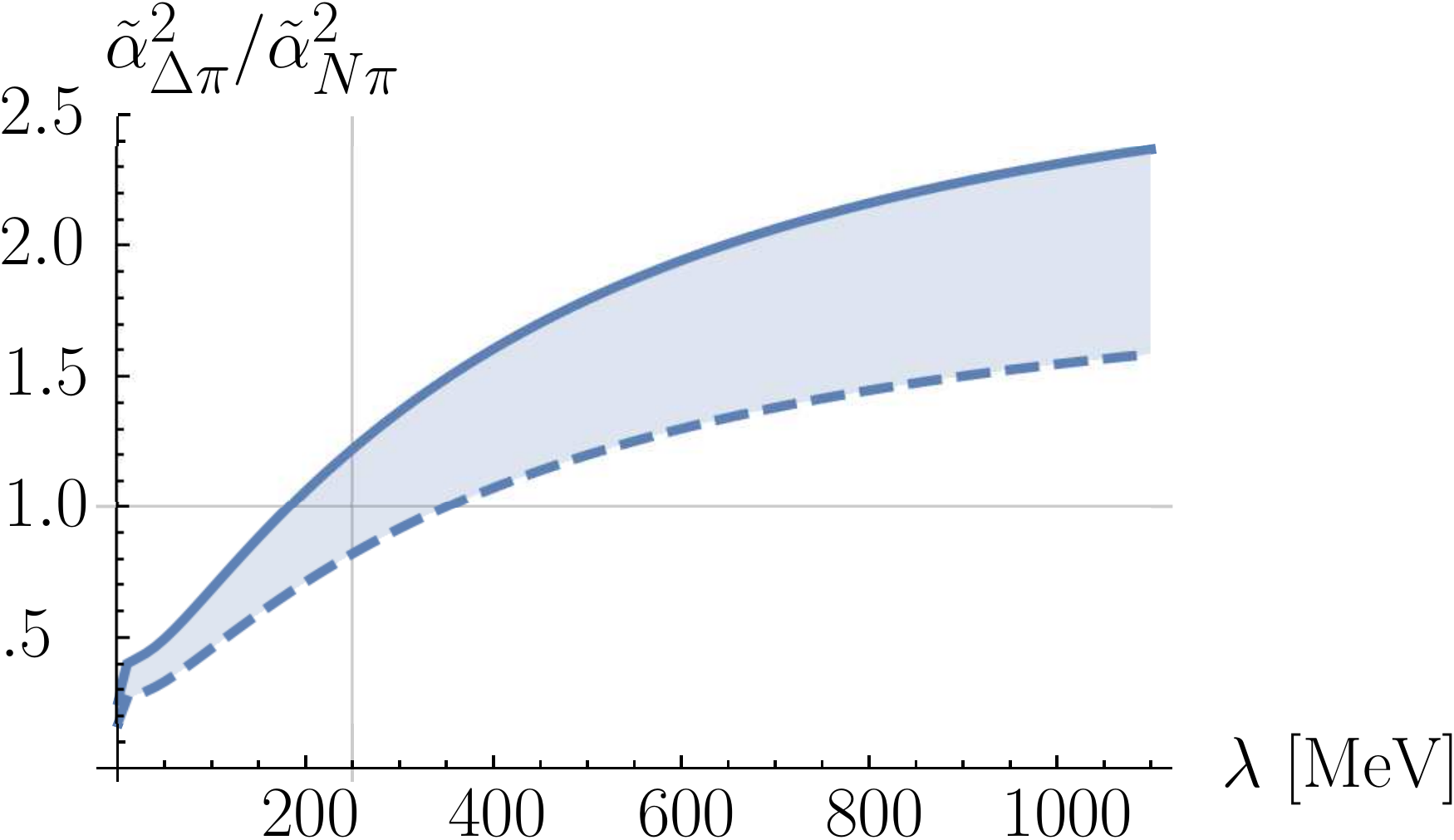}
    }
    \caption{The ratio $\tilde{\alpha}_{\Delta\pi}^2/\tilde{\alpha}_{N\pi}^2$ (a) Using the form factor $\phi_1$. (b) Using the form factor $\phi_2$. The bands are due to a variation in $h_A$: The solid (dashed) curves are for the larger $h_A^\text{max} = 2.937$ (smaller $h_A^\text{min} = 2.403$) value of $h_A$. The crossbars indicate when $\tilde{\alpha}_{\Delta\pi}^2/\tilde{\alpha}_{N\pi}^2 = 1$ which occurs at $(\Lambda,\lambda) \approx (750,250)$ MeV.}\label{Fig: ratio} 
\end{figure*}

It is also interesting to study the two dominant fluctuations, $N\pi$ and $\Delta\pi$, and their relative strengths. This was already presented in Figure \ref{Fig: probratios} (upper left corner). However, in Figure \ref{Fig: ratioLowUp1Plot} and \ref{Fig: ratioLowUp2Plot}, we show $\tilde{\alpha}_{\Delta\pi}^2/\tilde{\alpha}_{N\pi}^2$ using $\phi_1$ and $\phi_2$ respectively. We also include the effect a variation in the decuplet coupling $h_A$ has on this ratio. It is clear that the $\Delta\pi$ contribution is significant. For $\Lambda=750 ~\text{MeV}$ ($\lambda = 250$ MeV) it is of equal importance as the $N\pi$ term as indicated by the crossbars in the figure. 
Obviously depending on the exact value one uses for the coupling $h_A$ the $\Delta\pi$ term can become significant for even smaller (or larger) cut-off values as indicated by the band in Figure \ref{Fig: ratio} where the solid (dashed) curve corresponds to using $h_A^\text{max}$ ($h_A^\text{min}$). 

This observation of the magnitude of the $\Delta\pi$ term has implications for the interpretation of e.g.\ the $\bar{d}/\bar{u}$ asymmetry observed in the proton sea \cite{Towell:2001nh}. The $P\pi^0$ fluctuation is symmetric in $\bar{d}$ and $\bar{u}$ and cannot contribute to the asymmetry. While the $n\pi^+$ fluctuation can and indeed does contribute significantly to the ratio $\bar{d}/\bar{u}$, it does so too much to agree with experiment. Within the formalism presented here and in \cite{AESLGIHGasymm}, adding also the $\Delta\pi$ fluctuation to the cloud can explain the asymmetry in a natural way and to a quite good degree. The ratio $\bar{d}/\bar{u}$ is lowered due to the $\bar{u}$ in the $\pi^-$ in the $\Delta^{++}\pi^-$ fluctuation which, as shown in figures \ref{Fig: indprob1a} and \ref{Fig: indprob2a}, dominates over the $\Delta^0\pi^+$ which contains $\bar{d}$. 

Another interesting and elusive conundrum from a perturbative QCD point of view is that of a possible strange-quark asymmetry in the proton sea \cite{Bazarko:1994tt, Zeller:2002du, Alwall:2004rd, Aad:2012sb, Aad:2014xca}. In our model, we see that an asymmetry is present courtesy of the strange quarks in the fluctuations containing strangeness. 
This can be seen in figures \ref{Fig: eachContr1} and \ref{Fig: eachContr2} where the fluctuations containing strange quarks become non-negligible for reasonable cut-off values.\footnote{Apart from the $\Sigma K$ term which has a very small coupling to the proton. } We refer the interested reader to \cite{AESLGIHGasymm} for results concerning these issues. 

\begin{figure*}[!t]
	\centering
	\subfloat[]{\label{Fig: zall1}
	\includegraphics[width = 0.48\linewidth]{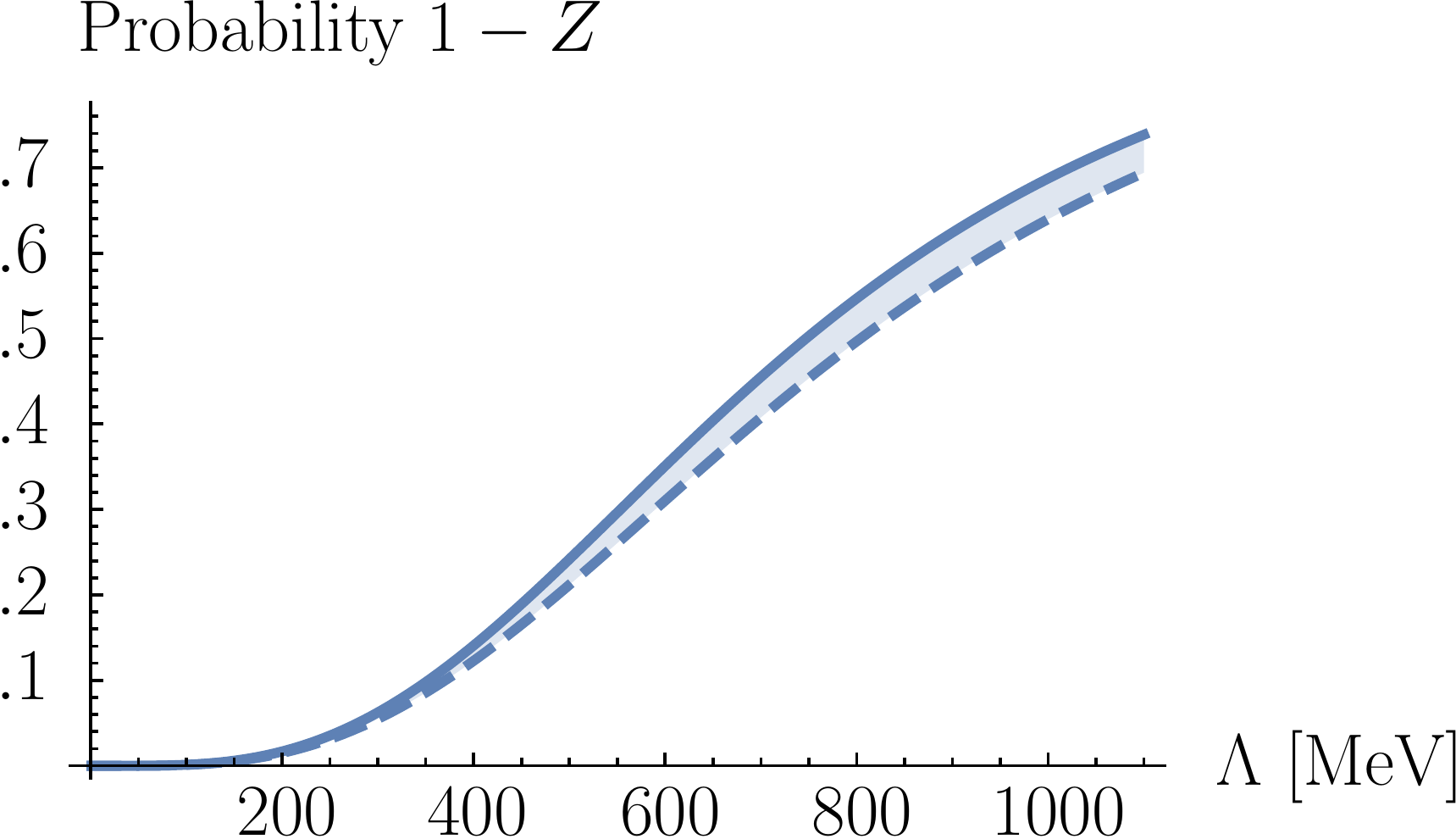}
	}
	\subfloat[]{\label{Fig: zall2}
	\includegraphics[width = 0.48\linewidth]{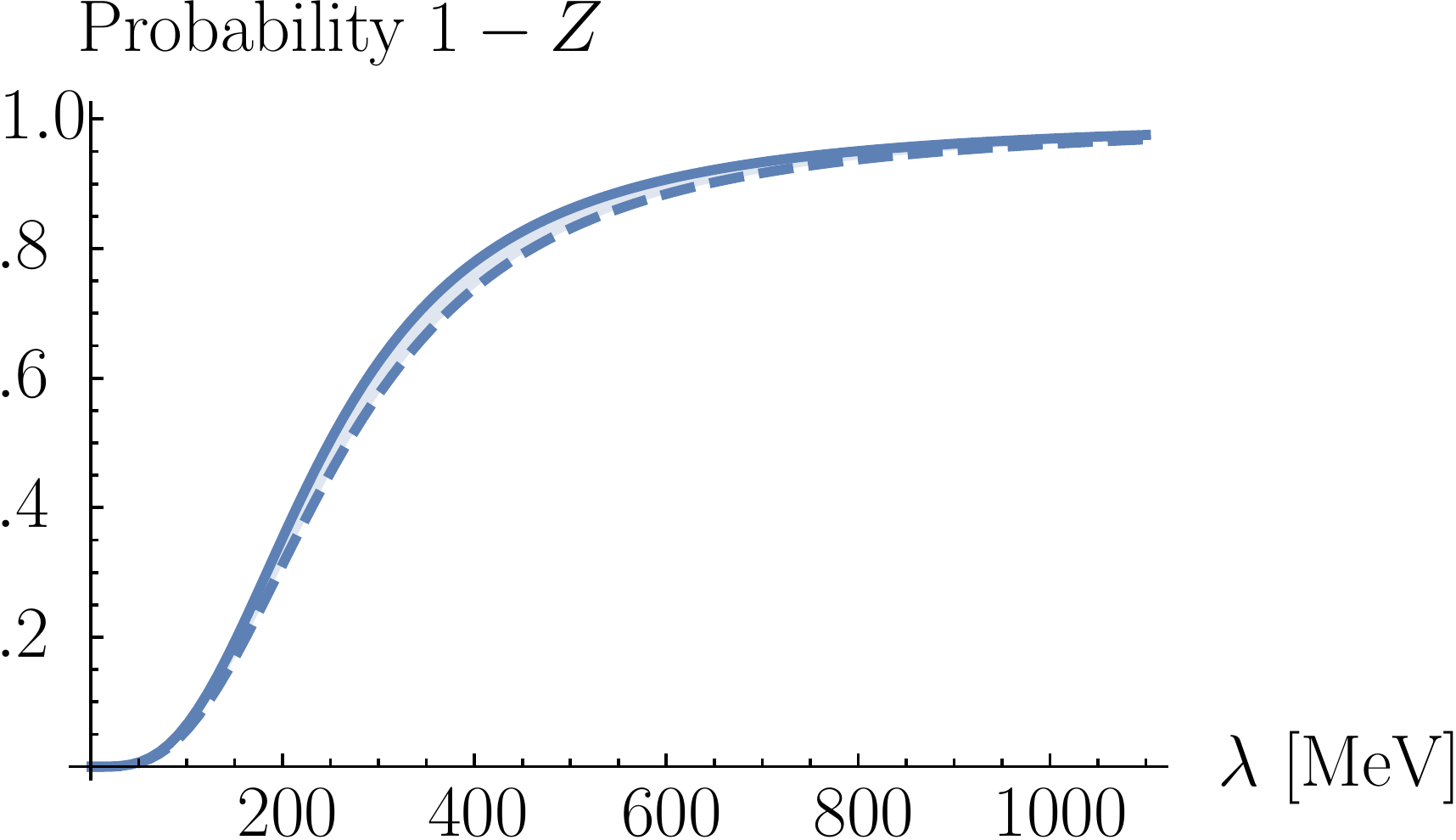}
	}
	\caption{The total fluctuation probability as a function of the cut-off parameter. (a) Using the form factor $\phi_1$. (b) Using the form factor $\phi_2$. The bands are due to a variation in $h_A$: The solid (dashed) curves are for the larger $h_A^\text{max} = 2.937$ (smaller $h_A^\text{min} = 2.403$) value of $h_A$. }
    \label{Fig: zAllVaryHa}
\end{figure*}

Finally we present the total fluctuation probability $1-Z$, given by Equation (\ref{E: totFlucProb}). This is shown in Figure \ref{Fig: zall1} and \ref{Fig: zall2} using $\phi_1$ and $\phi_2$ respectively. The band represents a variation in the decuplet coupling $h_A$. As seen a $10\%$ variation in the coupling $h_A$ results in a $5\%$ variation on the total fluctuation probability. This is to be contrasted to the results presented above where a $10\%$ variation in $h_A$ had a $10\%$ effect on the isospin-summed probabilities. 
\section{Discussion and Summary}\label{Sec: Conclusions}
We have derived explicit expressions for the coefficients of the hadronic fluctuation components of the proton wave function. We have done this using an effective approach that includes a form factor depending on a cut-off parameter. 

We have shown that for a reasonable form factor (a Gaussian) there is a near one-to-one relation between the cut-off parameter and the average momentum distribution in the fluctuations. We have shown how the hadronic fluctuation probabilities depend on the cut-off and we have shown their correlations. In particular we have shown that the two most probable fluctuations $N\pi$ and $\Delta\pi$ are of equal order for reasonable values of the cut-off parameter. 

The plot of the total fluctuation probability shown in Figure \ref{Fig: zAllVaryHa}, tells us that at realistic cut-off values, there is more than a $50\%$ probability that the proton is in a baryon-meson state rather than in a bare proton state. Therefore when the physical proton is probed, it is not unlikely that what one reads out in the end are not solely properties of the bare proton. This might lead to intriguing results. Indeed, there are plenty of experimental data (and theoretical models) that dismisses a simple Quark Model type static picture of the nucleon. These include the sea content of the proton \cite{Towell:2001nh,Bazarko:1994tt,Zeller:2002du,Alwall:2004rd,AESLGIHGasymm,AESLGIHGspin} and its spin structure \cite{Aidala:2012mv, Adolph:2016myg}. We refer the interested reader to \cite{AESLGIHGasymm} and \cite{AESLGIHGspin} for our new results on these issues. 
\section*{Acknowledgments}
The author thanks G.\ Ingelman and S.\ Leupold for valuable discussions and their careful reading of the manuscript. Fruitful discussions with C.\ Granados at the earlier stages of the work is gladly acknowledged. This work was supported by the Swedish Research Council (Vetenskapsr\r{a}det) under contract 621-2011-5107. 
\newpage
\appendix
\section{There is no contribution from the arc at infinity }\label{Sec: arcInfinity}
When doing the residue calculation in the complex $k_0$-plane, it is important to ensure that there is no contribution from the arc at infinity $\Gamma_\infty$ (cf.\ Figure \ref{Fig: Poles}). Therefore, in this section we take a closer look at the integrands of $\Sigma_{s}^i(p^2)$ and $F^i(p^2)$ with respect to $k_0$. We write the equations (\ref{E: sigmaOM}) \& (\ref{E: fom}) and (\ref{E: sigmadm}) \& (\ref{E: fdm}) on the form 
\eq{\label{E: sigmaOM2}
	\Sigma_s^{OM}(p^2) = -\ii |g_{OM}|^2 m_O\int \dbar^4 k |\phi|^2\frac{k^2}{\Delta_M \Delta_B}, 
}
\eq{\label{E: fom2}
	F^{OM}(p^2)= -\ii\frac{ |g_{OM}|^2}{p^2}
	\int \dbar^4k |\phi|^2\frac{k^2(k\cdot p +p^2)-2(k\cdot p)^2}{\Delta_M \Delta_B},
}
\eq{\label{E: sigmadm2}
	\Sigma_s^{DM}(p^2) = -\ii\frac{2 |g_{DM}|^2 m_D}{3}
	\int\dbar^4 k |\phi|^2 \frac{k^2p^2-(k\cdot p)^2}{\Delta_M \Delta_B}
}
and 
\eq{\label{E: fdm2}
	F^{DM}(p^2) = -\ii\frac{2 |g_{DM}|^2}{3p^2}
	\int\dbar^4 k |\phi|^2 \frac{(k\cdot p -p^2)((k\cdot p)^2-k^2 p^2)}{\Delta_M \Delta_B}, 
}
where 
\eq{\label{E: delta}
	\Delta_M &\equiv {k^2-m_M^2+\ii\epsilon}
	\\
	\Delta_B &\equiv {(p-k)^2-m_B^2+\ii\eta}. 
}
The relations (\ref{E: delta}) can be used in (\ref{E: sigmaOM2})\ -\ (\ref{E: fdm2}) in the form 
\eq{
	k^2  &= \Delta_M+m_M^2
	\\
	k\cdot p &= \frac{1}{2}(\Delta_M-\Delta_B+\mu^2),  
}
where we have defined $\mu^2 \equiv p^2-m_B^2+m_M^2$, which is independent of $k$. Then, the integrands can be written as 
\eq{
	\Sigma_s^{OM}: &\frac{k^2}{\Delta_M\Delta_B} = \frac{m_M^2}{\Delta_M\Delta_B}+\frac{1}{\Delta_B}. 
}
For $F^{OM}$ we get
\eq{\label{E: FomTemp}
	F^{OM}: \frac{k^2(k\cdot p +p^2)-2(k\cdot p)^2}{\Delta_M \Delta_B}
	&=
	\frac{-m_B^4+m_B^2 \left(m_M^2+2 p^2\right)+p^2
   \left(m_M^2-p^2\right)}{2 \Delta_M \Delta_B}
	\\
	&+\frac{p^2-m_B^2}{2 \Delta_M}+\frac{p^2+m_B^2}{2 \Delta_B}
}	
where we have dropped the term $k\cdot p/\Delta_M$ which integrates to zero. 

Next we consider $\Sigma_s^{DM}(p^2)$ given in Equation (\ref{E: sigmadm2}). 
One can expand the integrand as before 
\eq{
	\Sigma_s^{DM}: \frac{k^2p^2-(k\cdot p)^2}{\Delta_M \Delta_B} = 
	&\frac{-m_B^4+2 m_B^2 m_M^2+2 m_B^2 p^2-m_M^4-p^4}{2 \Delta_M\Delta_B
   }
   \\
   &
   +\frac{-m_B^2+m_M^2+p^2}{2 \Delta_M}+\frac{m_B^2-m_M^2+p^2}{2 \Delta_B}
   \\
   & -\frac{(k\cdot p)^2 }{\Delta_M\Delta_B} + \frac{\mu^2 k \cdot p }{\Delta_M\Delta_B}. 
}
But to show that there is no arc contribution, it is easiest to evaluate the original expression in the initial/final state proton's rest frame. Since then, one finds the integrand to be equal to 
\eq{
	\frac{-\vect{k}^2p_0^2}{\Delta_M\Delta_B}, 
}
which is fine with respect to the $k_0$ integration. 
The same arguments show that $F^{DM}$ is also fine with respect to $k_0$ integration. 

\bibliography{ProtonSelfEnergy.bib}
\end{document}